\documentclass[sort&compress,AMA,STIX1COL]{WileyNJD-v2}

\usepackage{bm}

\articletype{Research article}%

\received{xxx}
\revised{xxx}
\accepted{xxx}

\begin{document}

\title{{L}arge scale three-dimensional manufacturing tolerant stress-constrained topology optimization}

\author[1]{Gustavo Assis da Silva*}

\author[2]{Niels Aage}

\author[1]{Andr\'e Te\'ofilo Beck}

\author[2]{Ole Sigmund}

\authormark{DA SILVA \textsc{et al}}

\address[1]{\orgdiv{Department of Structural Engineering}, \orgname{S\~{a}o Carlos School of Engineering, University of S\~{a}o Paulo}, \orgaddress{\state{13.566-590, S\~{a}o Carlos, SP}, \country{Brazil}}}

\address[2]{\orgdiv{Department of Mechanical Engineering, Solid Mechanics}, \orgname{Technical University of Denmark}, \orgaddress{\state{Nils Koppels Alle, B. 404, 2800 Kgs. Lyngby}, \country{Denmark}}}

\corres{*Gustavo Assis da Silva, Department of Structural Engineering, S\~{a}o Carlos School of Engineering, University of S\~{a}o Paulo, 13.566-590, S\~{a}o Carlos, SP, Brazil. \\ \email{gustavoas@usp.br}}

\abstract[Summary]{In topology optimization, the treatment of stress constraints for very large scale problems has so far not been tractable due to the failure of robust agglomeration methods, i.e. their inability to accurately handle the locality of the stress constraints. This paper presents a three-dimensional design methodology that alleviates this shortcoming using both deterministic and robust problem formulations. The robust formulation, based on the three-field density projection approach, is extended to handle manufacturing uncertainty in three-dimensional stress-constrained problems. Several numerical examples are solved and further post-processed with body-fitted meshes using commercial software. The numerical investigations demonstrate that: (1) the employed solution approach based on the augmented Lagrangian method is able to handle large problems, with hundreds of millions of stress constraints; (2) if appropriate interpolation parameters are adopted, voxel-based (fixed grid) models can be used to compute von Mises stresses with excellent accuracy; and (3) in order to ensure manufacturing tolerance in three-dimensional stress-constrained topology optimization, a combination of double filtering and more than three realizations may be required.}


\maketitle

\section{Introduction}

Topology optimization is a widespread tool employed to achieve novel and high performance designs \citep{Aage_wing,CutFEM_2017_Maute,Alexandersen_2016,Wang_2018}. It consists in a material distribution method, employed to minimize/maximize an objective function (performance measure) while design constraints are satisfied \citep{BendsoeLivro,Sigmund2013}. An important, but challenging and less solved, application is stress-constrained topology optimization, especially in 3D.

Since the seminal paper by Duysinx and Bends\o e \cite{DuysinxBendsoe}, several works addressing topology optimization with stress constraints have been developed. Among them, there are works that focus on developing more efficient and accurate ways to handle the stress constraints \citep{IVE_Tensao,Troya2018,Guo_level_set,Sharma_Maute,Wang_Qian_2018,Nguyen_2020,Novotny_1,Bruggi_2016}, and works that focus on novel applications \citep{Moon2013,Collet2018,Helio_2019,Luo_2017_mambrane}, as the problem of compliant mechanisms with stress constraints \citep{DeLeon2015,Alexandre_mecanismos,Helio_2020,Artigo7,Artigo9}. Although not novel, stress-constrained topology optimization has been the subject of intensive research in the literature up to the present day.

Despite the remarkable achievements obtained so far, the current state-of-the-art in stress-constrained density-based topology optimization has not yet achieved the point of solving truly large scale three-dimensional problems within reasonable computational time. Besides the obvious need of more computational resources and use of parallel computing, one of the main difficulties that hinder this accomplishment is the local nature of the stress criterion, which implies a large number of stress constraints, and hence, potentially a large number of adjoint problems to be solved per optimization iteration \citep{Bruggi2012}. The number of stress constraints is usually equal to the number of elements in the finite element mesh if no special treatment is employed as, e.g., stress constraint aggregation techniques \citep{HolmbergTensao,DuysinxSigmund,ChauLe}. Although several techniques were developed to handle this issue, there are few papers addressing large scale three-dimensional stress-constrained topology optimization. To the authors' knowledge, the largest problem solved so far in the literature has been addressed by Leader et al. \cite{Leader_2019}, with $14$ million elements.

Although very important from the engineering design point of view, use of the stress constraint itself does not ensure robust and/or reliable designs. Real world engineering problems are often subjected to uncertainty in applied loads, material properties and manufacturing processes \citep{Melchers_Beck,BeyerRDO,Beck2012}. Most of the research in the field of stress-constrained topology optimization, however, is focused on deterministic problems. Although there are some recent works that propose formulations to handle uncertainty in stress-constrained topology optimization (see, e.g., \citep{RTO_Game_Theory,RBTO_Tensao,novotny_rbto_tensao,Holmberg_RTO_2,Artigo8,Artigo6,Artigo7,Artigo9}), these are applied to small scale problems, with less than $600$ thousand elements. The combination of stress constraints and manufacturing tolerances, despite its great importance for practical applications, is hence an unexplored topic in large scale three-dimensional topology optimization.

This paper addresses the topology optimization problem of volume minimization with stress constraints, and makes two main contributions to the state-of-the-art in the field:
\begin{enumerate}
\item The investigation and extension of existing robust formulations, based on the three-field density projection approach by Sigmund \cite{Sigmund2009} and Wang et al. \cite{Wang2011}, to handle manufacturing uncertainty in three-dimensional stress-constrained topology optimization;
\item The application of density-based topology optimization to address truly large scale three-dimensional stress-constrained problems (up to hundreds of millions of stress constraints).
\end{enumerate}

With this study, we extend the current methodologies for density-based topology optimization, in order to allow topology design of large scale three-dimensional structures which satisfy stress constraints and are robust with respect to manufacturing uncertainty. This paper can be seen as an extension of the work developed by Da Silva et al.\cite{Artigo6}, where two-dimensional volume minimization problems with stress constraints and manufacturing uncertainty are addressed. The extension, however, is not trivial, which makes it necessary to add additional discussions on computational and mechanical issues that arise from addressing the three-dimensional problems.

The paper is organized as follows. Section \ref{s2} presents the formulations addressed in this paper. Section \ref{s3} presents the solution method employed to solve the optimization problems in the manuscript. Section \ref{s4} presents several numerical results and important insights, and section \ref{s5} summarizes the main conclusions of this study. Appendices \ref{a1} and \ref{a2} present analytic studies on the one-dimensional filter and projection equations. Appendix \ref{a3} presents the sensitivity analysis. Appendix \ref{a4} presents additional insight on the solution procedure.

\section{Formulations}\label{s2}

In this paper, the density approach to topology optimization is employed: the topology optimization process is performed on a fixed finite element mesh (Eulerian approach), in which topology changes are allowed through variation of the relative densities \citep{Sigmund2013}. Each element is associated with a relative density $\overline{\rho}_{e}$, varying from $0$ (which represents void) to $1$ (which represents solid); these are updated through an iterative process, by use of a gradient-based algorithm, as described in section \ref{s3}.

Since the goal is to address truly large scale topology optimization problems, the PETSc-based topology optimization framework provided by Aage et al.\cite{Aage_PETSc} is employed as basis for the computational implementation. The structural problems are solved with the displacement-based finite element method for linear elasticity under static loads \citep{Bathe}, where 8-node linear brick elements are employed to discretize the design domains.

In this section, deterministic and robust formulations are presented. The deterministic formulation is well-known; it consists in the classical volume minimization problem with stress constraints considering the von Mises failure criterion \citep{ChauLe}, as presented in subsection \ref{s21}. The robust formulation, presented in subsection \ref{s22}, is novel, and represents an extension of the formulation developed by Da Silva et al.\cite{Artigo6}, which employs the three-field density projection approach by Wang et al. \cite{Wang2011}.

\subsection{Deterministic formulation}\label{s21}

The deterministic stress-constrained volume minimization problem in discrete form is written as
\begin{equation}
\begin{array}{lll}
\begin{array}{cc} \vspace{-12pt} \overset{\displaystyle \mathrm{Min.}}{^{\bm{\rho}}} \end{array} & V(\overline{\bm{\rho}}) = \sum_{e=1}^{N_{e}} V_{e} \overline{\rho}_{e} & \\ \\
\begin{array}{c} $ s. t.$ \end{array} & \frac{\sigma_{eq}^{(k)}(\overline{\bm{\rho}})}{\sigma_{y}} - 1 \leqslant 0 & \quad k=1,2,...,N_{k} \\
 & \mathbf{K}(\overline{\bm{\rho}})\mathbf{U}(\overline{\bm{\rho}}) = \mathbf{F} & \\ 
 & 0 \leqslant \rho_{e} \leqslant 1 & \quad e=1,2,...,N_{e}
\end{array},\label{Problema_otm_1}
\end{equation}
where $\bm{\rho} \in \mathbb{R}^{N_{e}}$ are the design variables of the optimization problem, $V(\overline{\bm{\rho}})$ is the structural volume, $\overline{\bm{\rho}} \in \mathbb{R}^{N_{e}}$ are the relative densities, $N_{e}$ is the number of finite elements in the mesh, $V_{e}$ is the structural volume of element $e$, $\sigma_{eq}^{(k)}(\overline{\bm{\rho}})$ is the von Mises equivalent stress at point $k$, $\sigma_{y}$ is the yield stress, $N_{k}$ is the number of points where the von Mises stress is computed, $\mathbf{K}(\overline{\bm{\rho}})$ is the global stiffness matrix, $\mathbf{U}(\overline{\bm{\rho}})$ is the global displacement vector and $\mathbf{F}$ is the global load vector. The local stiffness matrix of element $e$ is computed with the Solid Isotropic Material with Penalization (SIMP) scheme \citep{BendsoeLivro}, as $\mathbf{k}_{e}(\overline{\rho}_{e}) = \left( \rho_{min} + \left( 1 - \rho_{min} \right)\overline{\rho}_{e}^{p} \right) \mathbf{k}_{e}^{0}$, where $\mathbf{k}_{e}^{0}$ is the $e$-th local stiffness matrix of base material, $\rho_{min} = 10^{-9}$ is adopted to avoid singularity issues when solving for equilibrium, and $p=3$ is used as penalization factor.

Relative densities are related to design variables through a density filter with threshold projection \citep{Wang2011}. In this paper, filtered densities, given by $\tilde{\bm{\rho}}$, are obtained implicitly by solving a Partial Differential Equation (PDE) with homogeneous Neumann boundary conditions \citep{Lazarov_PDE}, given by
\begin{equation}
-R_{PDE}^2 \nabla^2 \tilde{\rho} + \tilde{\rho} = \rho, \quad \frac{\partial \tilde{\rho}}{\partial \mathbf{n}}=0,\label{filtro_PDE}
\end{equation}
where $R_{PDE} \geqslant 0$ controls the length scale.

The threshold projection, which relates relative densities to filtered densities \citep{Wang2011}, is given by a smoothed Heaviside function
\begin{equation}
\overline{\rho}_{e} = \frac{\tanh{(\beta \eta)} + \tanh{(\beta (\tilde{\rho}_{e} - \eta))}}{\tanh{(\beta \eta)} + \tanh{(\beta (1 - \eta))}},\label{projecao_Heaviside}
\end{equation}
where $\tilde{\rho}_{e}$ is the filtered relative density of element $e$, $\eta \in (0,1)$ is a user-defined parameter that controls the inflection point of the threshold projection: for $\eta \rightarrow 0$ we have dilation behavior, for $\eta \rightarrow 1$ we have erosion behavior, and for $\eta = 0.5$ we approach volume preserving behavior \citep{Wang2011}; and $\beta > 0$ controls the sharpness of the projection: the larger the value of $\beta$, the smaller the amount of intermediate material in the topology.

Lazarov and Sigmund \cite{Lazarov_PDE} propose an approximate relation between the length scales for classical and PDE filters, given by $R = 2 \sqrt{3} R_{PDE}$, where $R$ is the support domain of the classical filter with linear hat function. As observed by Lazarov and Sigmund\cite{Lazarov_PDE}, however, the PDE filter compared to the classical filter with linear hat function has higher weights for points close to the point where the filtering is performed and hence, the obtained results are more black and white for equivalent length scales. This phenomenon directly affects the stress-based formulation employed in this paper, since the amount of intermediate material between solid and void phases should be large enough to alleviate the undesirable effects related to the jagged boundaries, as observed by Da Silva et al. \cite{Artigo6}.

In order to ensure a smooth transition boundary of length equal to the side of a square element $l_{e}$ (or cubic element, in three-dimensions), between solid and void phases, Da Silva et al. \cite{Artigo6} defined an upper bound for $\beta$, to be used in Equation \eqref{projecao_Heaviside}, given by $\beta_{lim} = \frac{2 R}{l_{e}}$. In  Appendix \ref{a1}, however, we demonstrate that a different upper bound should be employed when using the PDE filter. This value is given by $\beta_{lim}^{PDE} = \frac{2 R}{l_{e} \sqrt{3}}$. This is in agreement with the behavior observed by Lazarov and Sigmund \cite{Lazarov_PDE}, in the sense that we have to use a smaller $\beta$ value after a PDE filtering operation to ensure the same gray-scale of a projection after classical linear filtering. In order to enjoy the benefits of stress accuracy and smooth stress behavior after uniform boundary variation, we thus employ a maximum value of $\beta_{max} \cong \beta_{lim}^{PDE} / 2$ in the optimization procedure.

The von Mises equivalent stress is computed based on Duysinx and Bends\o e \cite{DuysinxBendsoe}, and is written as
\begin{align}
\sigma_{eq}^{(k)}\left(\overline{\bm{\rho}}\right) & = f_{\sigma} \left( \overline{\rho}_{k} \right) \hat{\sigma}_{eq}^{(k)}\left(\overline{\bm{\rho}}\right) \nonumber \\
 & = f_{\sigma} \left( \overline{\rho}_{k} \right) \sqrt{\hat{\bm{\sigma}}_{k}^{T}\left(\overline{\bm{\rho}}\right) \mathbf{M} \hat{\bm{\sigma}}_{k}\left(\overline{\bm{\rho}}\right) + \sigma_{min}^{2}},\label{tensao_von_Mises_2}
\end{align}
where $f_{\sigma} \left( \overline{\rho}_{k} \right)$ is the stress interpolation function, $\hat{\sigma}_{eq}^{(k)}\left(\overline{\bm{\rho}}\right)$ is the solid von Mises stress at point $k$, $\hat{\bm{\sigma}}_{k}\left(\overline{\bm{\rho}}\right)$ is the solid stress vector at point $k$, $\sigma_{min} = 10^{-4} \sigma_{y}$ is a small value included in our implementations to ensure a positive von Mises equivalent stress when $\hat{\bm{\sigma}}_{k}^{T}\left(\overline{\bm{\rho}}\right) \mathbf{M} \hat{\bm{\sigma}}_{k}\left(\overline{\bm{\rho}}\right) \rightarrow 0$, thus avoiding numerical instabilities during the sensitivity analysis, and $\mathbf{M}$ is the standard operator matrix for calculating the von Mises stresses.

The solid stress vector is given by
\begin{equation}
\hat{\bm{\sigma}}_{k}\left(\overline{\bm{\rho}}\right) = \mathbf{C}^{0} \mathbf{B}_{k} \mathbf{u}_{k}(\overline{\bm{\rho}}),\label{tensao}
\end{equation}
where, $\mathbf{C}^{0}$ is the constitutive matrix of the base material, $\mathbf{B}_{k}$ is the strain-displacement transformation matrix evaluated at point $k$ and $\mathbf{u}_{k}(\overline{\bm{\rho}})$ is the local displacement vector of the element which contains point $k$.

The stress interpolation function, $f_{\sigma} \left( \overline{\rho}_{k} \right)$, must be properly chosen to avoid the singularity phenomenon \citep{ChauLe}. In this paper, the $\varepsilon$-relaxed approach is employed \citep{DuysinxSigmund,ChengGuo}, with $f_{\sigma} \left( \overline{\rho}_{k} \right) = \frac{\overline{\rho}_{k}}{\varepsilon\left( 1-\overline{\rho}_{k} \right)+\overline{\rho}_{k}}$. As demonstrated by Da Silva et al. \cite{Artigo6}, parameter $\varepsilon$ plays an important role in stress accuracy at the interface between solid and void regions. In this paper, we use $\varepsilon = 0.2$, following Da Silva et al. \cite{Artigo6}, since this is a good choice when associated with $\beta_{max} \cong \beta_{lim}^{PDE} /2$.

\subsection{Robust formulation}\label{s22}

The proposed robust formulation is based on the three-field density projection approach by Wang et al. \cite{Wang2011}, since the goal is to achieve optimized topologies that are insensitive to uniform boundary variations. The three-field approach considers three sets of relative densities during the topology optimization process: $\overline{\bm{\rho}}^{(d)}$, $\overline{\bm{\rho}}^{(i)}$ and $\overline{\bm{\rho}}^{(e)}$; representing dilated, intermediate and eroded topologies, respectively. These are obtained for different values of $\eta$ in Equation \eqref{projecao_Heaviside}, following the relation: $\eta_{d} < \eta_{i} < \eta_{e}$. In this formulation, eroded and dilated topologies represent extreme manufacturing errors. By taking these additional fields into account during the optimization process, it is expected to achieve a robust (intermediate) topology that is tolerant to manufacturing errors with the underlying assumption that the entire boundary may be eroded or dilated by the same amount.

The immediate application of the three-field robust approach, however, turns out not to be effective in ensuring manufacturing error tolerant topologies in three-dimensional stress-based design, since these have shown to be more sensitive to uniform boundary variations. This observation leads to the generalization of the standard three-field formulation by:
\begin{enumerate}
\item Allowing more density field realizations;
\item Applying the double filter approach, by Christiansen et al. \cite{Christiansen2015}, to address the problem, instead of the single filter approach as presented so far in the manuscript.
\end{enumerate}

The proposed robust formulation is written as
\begin{equation}
\begin{array}{lll}
\begin{array}{cc} \vspace{-12pt} \overset{\displaystyle \mathrm{Min.}}{^{\bm{\rho}}} \end{array} & V\left(\overline{\overline{\bm{\rho}}}^{(1)}\right) = \sum_{e=1}^{N_{e}} V_{e} \overline{\overline{\rho}}_{e}^{(1)} & \\ \\
\begin{array}{c} $ s. t.$ \end{array} & \frac{\sigma_{eq}^{(k)}\left(\overline{\overline{\bm{\rho}}}^{(j)}\right)}{\sigma_{y}} - 1 \leqslant 0 & \quad j=1,2,...,N_{j} \; $  and  $ \; k=1,2,...,N_{k} \\
 & \mathbf{K}\left(\overline{\overline{\bm{\rho}}}^{(j)}\right)\mathbf{U}\left(\overline{\overline{\bm{\rho}}}^{(j)}\right) = \mathbf{F} & \quad j=1,2,...,N_{j} \\ 
 & 0 \leqslant \rho_{e} \leqslant 1 & \quad e=1,2,...,N_{e}
\end{array},\label{Problema_otm_3}
\end{equation}
where index $j$ refers to the $j$-th density realization and $N_{j}$ is the total number of fields of relative densities. Each field of relative densities $\overline{\overline{\bm{\rho}}}^{(j)}$ is associated with a parameter $\eta_{j}$ in Equation \eqref{projecao_Heaviside}. We consider $\eta_{1} < \eta_{2} < ... < \eta_{N_{j}-1} < \eta_{N_{j}}$, meaning that index $j=1$ refers to the most dilated topology, whereas index $j=N_{j}$ refers to the most eroded topology. Note that the volume of the dilated topology is minimized in this formulation, in agreement with the previous approach by Da Silva et al. \cite{Artigo6}. Using the volume of the dilated topology as objective works at avoiding the undesirable numerical instabilities that arise when handling the volume of the intermediate topology directly in the formulation \citep{Sigmund2009,Wang2011}.

The double filter approach consists in the single filter procedure with $N_{j}$ fields of relative densities, and additional filter and projection steps. It is given by:
\begin{enumerate}
\item The filtering operation, Equation \eqref{filtro_PDE}, is applied on the design variables, given by $\bm{\rho}$, considering a filter radius of $2 R_{PDE}$, to obtain the first level filtered densities, given by $\tilde{\bm{\rho}}$;
\item The projection step, Equation \eqref{projecao_Heaviside}, is applied on the first level filtered densities, considering $\eta = \min \left\lbrace \eta_{1}, \eta_{2}, ... , \eta_{N_{j}-1}, \eta_{N_{j}} \right\rbrace$ and $2 \beta$, to obtain the first level projected densities, given by $\overline{\bm{\rho}}$;
\item The filtering operation is applied on the first level projected densities, for a filter radius of $R_{PDE}$, to obtain the second level filtered densities, given by $\tilde{\overline{\bm{\rho}}}$;
\item The projection step is applied on the second level filtered densities, considering $\eta_{1}, \eta_{2}, ... , \eta_{N_{j}-1}, \eta_{N_{j}}$ (as many fields as necessary) and $\beta$, to obtain the second level projected densities, given by $\overline{\overline{\bm{\rho}}}^{(1)} , \overline{\overline{\bm{\rho}}}^{(2)} , ... , \overline{\overline{\bm{\rho}}}^{(N_{j}-1)} , \overline{\overline{\bm{\rho}}}^{(N_{j})}$. These are the actual fields of physical relative densities.
\end{enumerate}

Christiansen et al. \cite{Christiansen2015} proposed the robust double filter approach to address highly sensitive topology optimization problems. During our numerical investigations on three-dimensional stress-based design, we observed that use of the robust single filter approach may lead to unusable optimized designs, with different topologies for different projection levels. The robust double filter procedure turns out to resolve the problem, providing identical eroded, intermediate and dilated topologies, and it comes at virtually no extra cost compared to the single filter procedure.

Use of the double filter in stress-constrained topology optimization is not novel. In Da Silva et al. \cite{Artigo9}, the robust double filter approach is employed to address the path-generating problem subjected to stress constraints and manufacturing uncertainty. In that case, the standard single filter procedure was not sufficient to remove all gray areas and ensure manufacturing tolerant designs.

Note that the deterministic formulation, Equation \eqref{Problema_otm_1}, is the particularization of the proposed robust formulation, Equation \eqref{Problema_otm_3}, for $N_{j} = 1$ and $\overline{\overline{\bm{\rho}}} = \overline{\bm{\rho}}$. The same applies for the robust three-field approach by Da Silva et al. \cite{Artigo6}, for $N_{j} = 3$ and $\overline{\overline{\bm{\rho}}} = \overline{\bm{\rho}}$.

\section{Solution approach}\label{s3}

In this paper, where truly large scale problems are addressed, the number of stress constraints becomes extremely large. For this purpose, we employ the augmented Lagrangian method, which turns out to be a valid approach to solve problems with up to hundreds of millions of stress constraints. This method has been shown to be a valid alternative to the aggregation techniques often employed to handle the large number of stress constraints in the formulation. Introduced by Pereira et al. \cite{Pereira2004} in the field of stress-constrained topology optimization, the augmented Lagrangian method has been employed to address density-based  \citep{Artigo2,Artigo3,Artigo6,Artigo7,Artigo8,Artigo9,Alexandre_mecanismos,Fancello_2006,Fancello_Tensao} and level-set based \citep{Fancello_Level_set1,Fancello_Level_set2,Helio_2019,Helio_2020} problems with stress constraints.

In this work, the augmented Lagrangian formulation by Birgin and Mart\'{i}nez \cite{Martinez} is employed. The augmented Lagrangian method replaces the original constrained optimization problem by a sequence of bound constrained optimization subproblems. The augmented Lagrangian function is given by
\begin{equation}
L\left(\bm{\rho},\bm{\mu},r\right) = \frac{N_{k}}{\sum_{e=1}^{N_{e}}V_{e}} V\left(\overline{\overline{\bm{\rho}}}^{(1)}\right) + \frac{r}{2}\sum_{k=1}^{N_{k}} \sum_{j=1}^{N_{j}} \left\langle \frac{\mu_{k}^{(j)}}{r} + \frac{\sigma_{eq}^{(k)}\left(\overline{\overline{\bm{\rho}}}^{(j)}\right)}{\sigma_{y}} - 1 \right\rangle^{2},\label{LA_robusto}
\end{equation}
where $\langle \cdot \rangle = \max(0,\cdot)$, $r$ is the penalization parameter, and $\mu_{k}^{(j)}$ is the Lagrange multiplier associated with $k$-th stress constraint and $j$-th relative density field. Note that all stress constraints are included in the augmented Lagrangian function and that objective function (structural volume) is weighted by constant $\frac{N_{k}}{\sum_{e=1}^{N_{e}}V_{e}}$ for the purpose of normalization. Besides, we normalize the initial, $r_{0} = 0.01$, and final, $r_{max} = 10000$, penalization parameters (to be employed during the optimization procedure), as $\frac{r_{0}}{N_{j}}$ and $\frac{r_{max}}{N_{j}}$, respectively.


The optimization subproblems are given by
\begin{equation}
\begin{array}{lll}
\begin{array}{cc} \vspace{-12pt} \overset{\displaystyle \mathrm{Min.}}{^{\bm{\rho}}} \end{array} & L\left(\bm{\rho},\bm{\mu},r\right) & \\  \\
\begin{array}{c} $ s. t.$ \end{array} & \mathbf{K}\left(\overline{\overline{\bm{\rho}}}^{(j)}\right)\mathbf{U}\left(\overline{\overline{\bm{\rho}}}^{(j)}\right) = \mathbf{F} & \quad j=1,2,...,N_{j} \\ 
 & 0 \leqslant \rho_{e} \leqslant 1 & \quad e=1,2,...,N_{e}
\end{array},\label{Problema_otm_4}
\end{equation}
which are solved with the steepest descent method with move limits (Appendix \ref{a4}).

We adopt an approximate iterative procedure, in the sense that optimization subproblems are not solved strictly up to a prescribed tolerance. Instead, we update the Lagrange multipliers every $20$ iterations, and the penalization parameter every $60$ iterations. Despite this simplification, we have achieved good results within a reasonable number of iterations.

Lagrange multipliers and penalization parameter are updated by
\begin{equation}
\mu_{k}^{(j)} \leftarrow \left\langle r \left(\frac{\sigma_{eq}^{(k)}\left(\overline{\overline{\bm{\rho}}}^{(j)}\right)}{\sigma_{y}} - 1\right) + \mu_{k}^{(j)} \right\rangle,\label{Multiplicador_de_Lagrange}
\end{equation}
\begin{equation}
r \leftarrow \min\left(\gamma \ r,r_{max}\right),\label{Parametro_de_penalizacao}
\end{equation}
where $\gamma = 10$ is an update parameter.

The augmented Lagrangian method does not take into account the $\beta$-continuation scheme \citep{Guest_Heaviside,Sigmund-2007}, often employed in density-based topology optimization to ensure numerically stable procedure. In the current implementation, the value of $\beta$ starts very small, and it is increased every $100$ iterations by the following relation: $\beta \leftarrow \min\left(2 \ \beta,\beta_{max}\right)$.

\section{Results and discussion}\label{s4}

Several numerical examples are addressed to demonstrate the applicability of the proposed approach, Equation \eqref{Problema_otm_3}. The systems of linear equations resulting from the state and adjoint problems are solved with the Galerkin projection geometric multigrid preconditioned flexible-GMRES \citep{Aage_PETSc}. A W-cycle framework with four multigrid levels is employed, with four FGMRES/SOR smoothing steps per level and a relative convergence tolerance of $10^{-5}$. The coarse level problem is solved with FGMRES/SOR to $10^{-8}$ or a maximum of $30$ iterations. The optimization problems are solved on the DTU Sophia cluster using $32$ nodes each containing 2 AMD EPYC 7351 16 core processors and 128 GB RAM memory, when truly large problems are addressed ($> 100$ million elements), and only $2$ nodes otherwise.

\begin{figure}[ht!]
\centering
\includegraphics[width=0.99\textwidth]{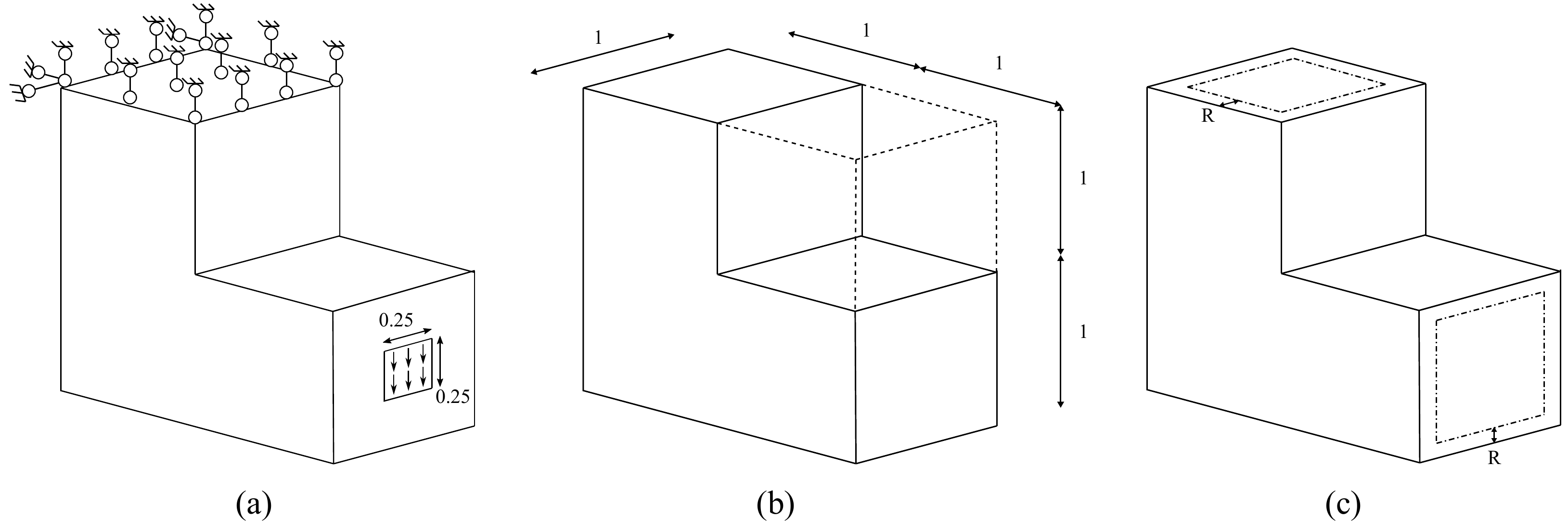}
\caption{(a) L-shaped problem; (b) Full domain considered for finite element discretization; non-design field is represented with dashed lines and design field is represented with solid lines; (c) Internal boundary padding scheme; it surrounds the entire boundary of the design domain, except at support and load regions.}
\label{L1}
\end{figure}

A three-dimensional version of the L-shaped design problem, Figure \ref{L1} (a), is adopted as main example. Boundary conditions: roller support on the upper surface and vertical distributed load on the right surface. Three additional degrees of freedom are constrained to avoid rigid body motion, Figure \ref{L1} (a). In order to address this problem, we constructed a structured finite element mesh of physical dimension $2 \times 1 \times 2$, thus encompassing the L-shaped design domain, Figure \ref{L1} (b). Note that design variables in the non-design region are kept equal to zero during the whole optimization procedure. Figure \ref{L1} (c) illustrates an internal padding region of width equal to $R$ (or $2R$ when the double filter is considered, to encompass the first level filtered densities) that surrounds the entire boundary of the design domain, except at support and load regions. This region is filled with design variables equal to zero, and the gradient components at these points are set to zero. The internal padding approach employed in this manuscript helps to alleviate possible boundary effects that may occur due to filtering, like the exterior padding approaches by Clausen and Andreassen \cite{Clausen2017}, but the internal padding is chosen since it is much simpler to implement. Although design variables are zero in the domains, filtered and projected densities may be different from zero.

Input data are: Young's modulus $1$, Poisson's ratio $0.3$, yield stress $\sigma_{y} = 50$, and the total applied load is set to unity and distributed on a square-shaped region of dimensions $0.25 \times 0.25$. Von Mises equivalent stresses are computed at the centroid of each 8-node linear brick element. Regarding the optimization procedure, we consider an initial value for the design variables of $\bm{\rho}=\mathbf{1}$, except at non-design and padding regions. The other parameters that were not mentioned in this paragraph were given in sections \ref{s2} and \ref{s3}, with the exception of the filter radius $R$, and parameters $\eta$ and $\beta$, associated with the threshold projection; these are presented in each subsection.


The iterative procedure is performed until $\beta$ completes $100$ iterations at the maximum projection level, and the stress constraints are satisfied considering a tolerance of $5\%$ regarding the yield stress. Although this convergence criterion is not conservative, we were able to obtain good results within reasonable number of iterations in all cases.

\subsection{Deterministic \texorpdfstring{$\times$}{Lg} Robust}\label{s41}

In this subsection, we solve the L-shaped problem with the deterministic and the robust double filter approaches, Equations \eqref{Problema_otm_1} and \eqref{Problema_otm_3}, respectively. The design domain is discretized with a mesh of $320 \times 160 \times 320 \cong 16.4$ million elements. A filter radius of $R = 0.04$ is employed. The edge of a cubic element, $l_{e} = 1/160 = 0.00625$, is used to compute $\beta_{lim}^{PDE} = \frac{2 R}{l_{e} \sqrt{3}} \cong 7.390$. Following Da Silva et al. \cite{Artigo6}, we use $\beta_{max} = 3.695 \cong \beta_{lim}^{PDE}/2$. We start with the first $\beta$ smaller than $0.25$ when successively dividing $\beta_{max}$ by two, i.e., $\beta = \frac{3.695}{2^{4}}$; in this case, it takes $4$ updates (one every $100$ iterations) to reach $\beta_{max}$, and then, further $100$ iterations at the last level, resulting in a minimum of $500$ iterations before termination. Stopping criterion: if the stress constraints are satisfied at $500$ iterations, the procedure is completed; otherwise, the iterative procedure continues until these are satisfied.

\begin{figure}[ht!]
\centering
\includegraphics[width=0.99\textwidth]{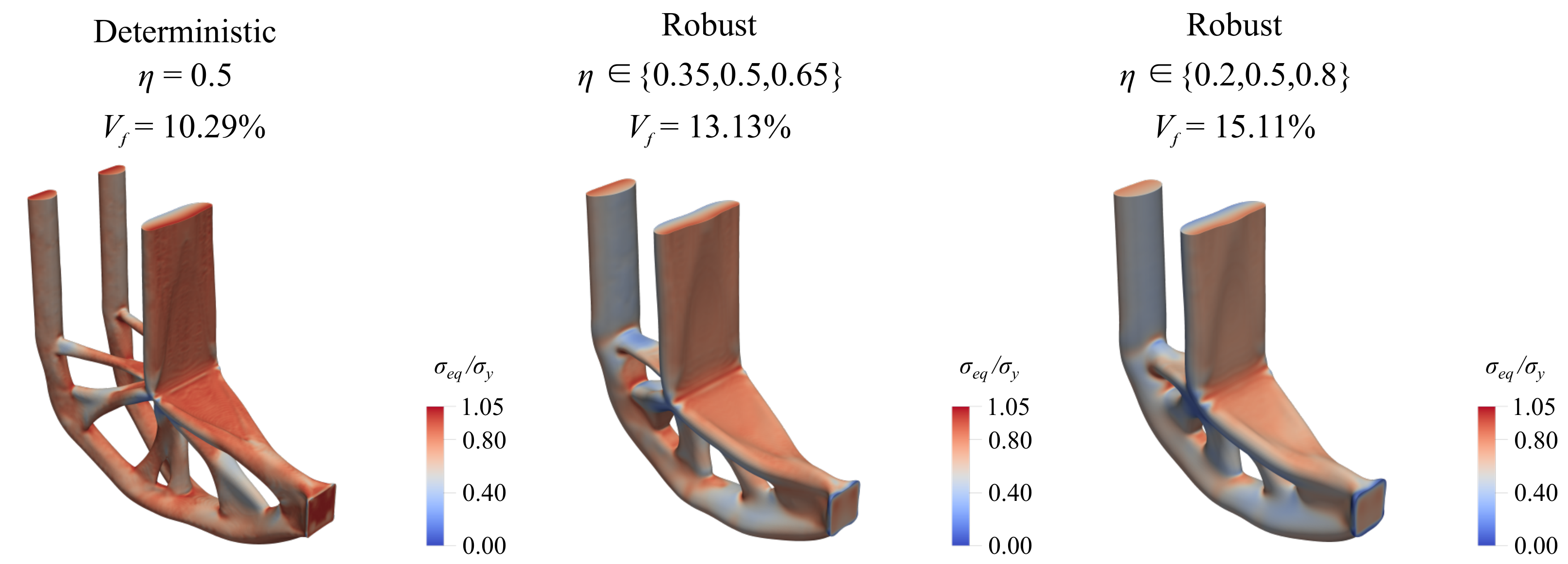}
\caption{Optimized intermediate topologies for different situations regarding parameter $\eta$. Color scales indicate normalized von Mises stresses.}
\label{L1_topologies}
\end{figure}

The deterministic problem is solved for $\eta = 0.5$. The robust problems are solved for $N_{j} = 3$ (three fields of relative densities) and two situations regarding parameter $\eta$, as follows: (1) $\eta \in \left\lbrace 0.35 , 0.5 , 0.65 \right\rbrace$; and (2) $\eta \in \left\lbrace 0.2 , 0.5 , 0.8 \right\rbrace$. Figure \ref{L1_topologies} illustrates optimized intermediate topologies (i.e., for $\eta = 0.5$) with respect to von Mises stress distributions and volume fractions, $V_{f}$. Visualization is performed with ParaView \citep{paraview}; the Iso Volume filter is applied for surface smoothing.

\begin{figure}[ht!]
\centering
\includegraphics[width=0.99\textwidth]{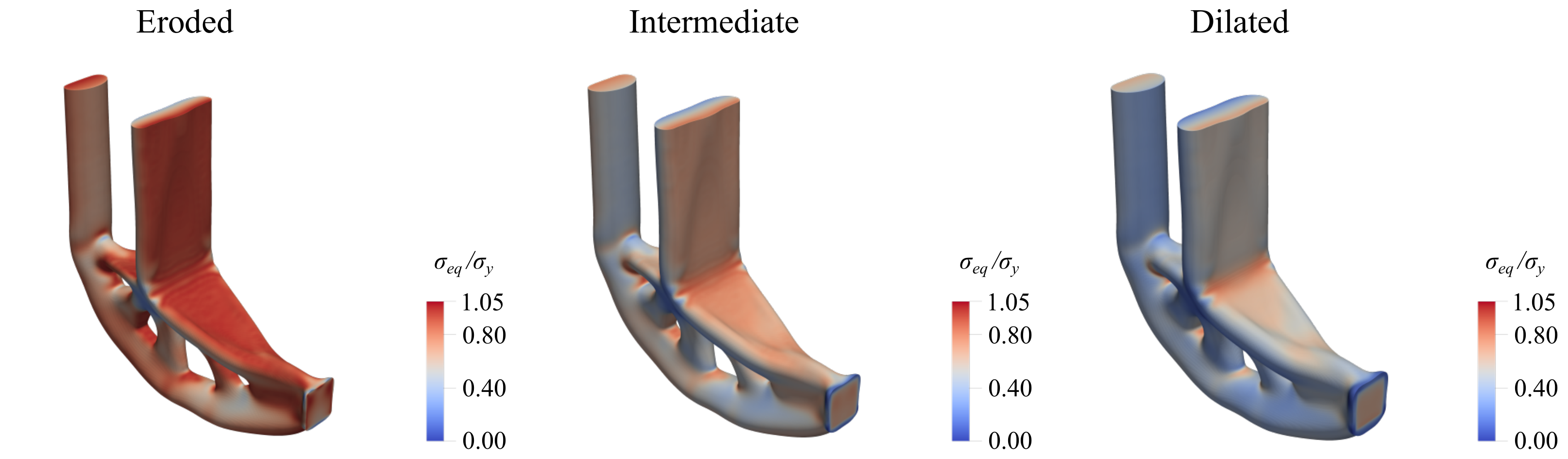}
\caption{Eroded, intermediate and dilated optimized topologies of robust result for $\eta \in \left\lbrace 0.2 , 0.5 , 0.8 \right\rbrace$.}
\label{L1_DF_eta03_topologies}
\end{figure}

Analyzing Figure \ref{L1_topologies}, one observes that optimized topologies obtained by deterministic and robust approaches are different, with the latter presenting less structural details. When analyzing the volume fractions: the deterministic solution has the smallest volume fraction among all solutions presented. This is not surprising, since the deterministic design does not have to fulfill any manufacturing requirement. Figure \ref{L1_DF_eta03_topologies} shows eroded, intermediate and dilated topologies for the case with $\eta \in \left\lbrace 0.2 , 0.5 , 0.8 \right\rbrace$. The same behavior described by Da Silva et al. \cite{Artigo6} is observed in this case: the eroded topology is the most stressed one, although the other two topologies have some highly stressed regions as well.

\begin{figure}[ht!]
\centering
\includegraphics[width=0.99\textwidth]{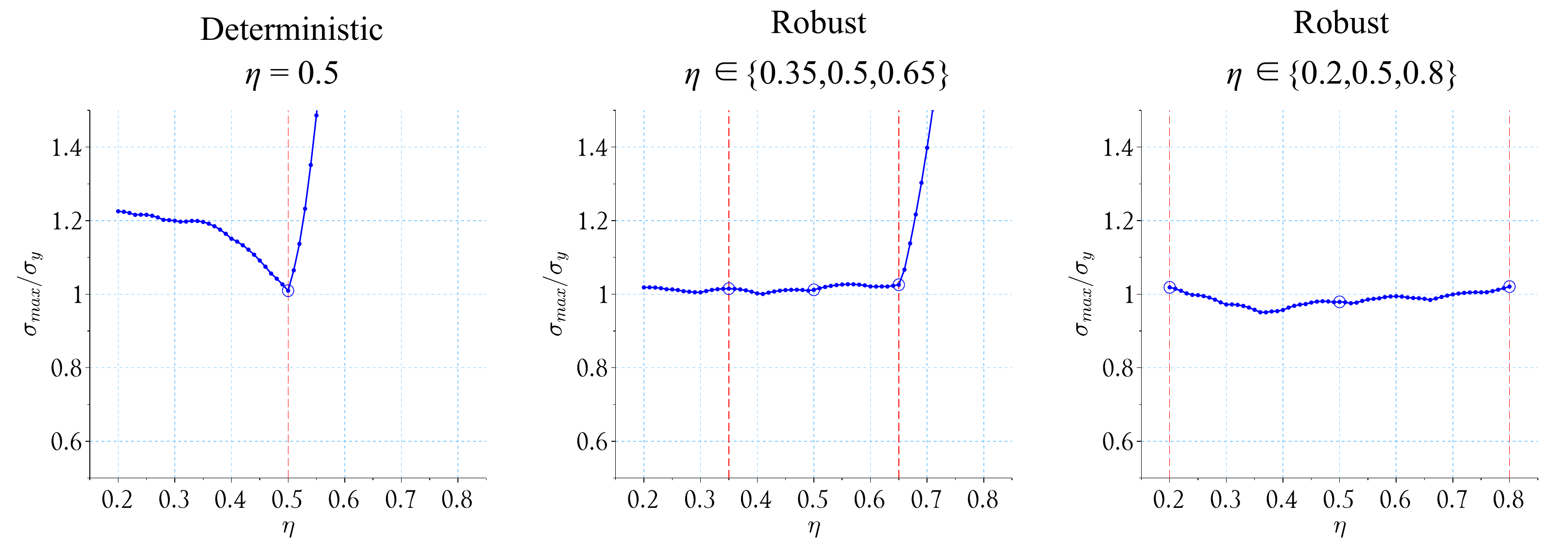}
\caption{Post-processing graphs for maximum normalized von Mises stresses of topologies in Figure \ref{L1_topologies}. Vertical dashed lines (in red) indicate the $\eta$ range considered during optimization. Relative density fields considered during optimization are indicated by circles.}
\label{L1_stresses}
\end{figure}

In order to check for manufacturing tolerance, the maximum von Mises stress, $\sigma_{max}$, is plotted for several values of $\eta$ between $0.2$ and $0.8$, using the voxel-based post-processing scheme by Da Silva et al. \cite{Artigo6}, Figure \ref{L1_stresses}. Analyzing the deterministic stress graph, one can verify stress feasibility for $\eta = 0.5$ only, i.e., the deterministic result is not robust at all, since erosion or dilation operations lead to structures that present maximum von Mises stresses that are much larger than the yield stress. When analyzing the cases for $\eta \in \left\lbrace 0.35 , 0.5 , 0.65 \right\rbrace$ and $\eta \in \left\lbrace 0.2 , 0.5 , 0.8 \right\rbrace$, one can verify stress feasibility for wider ranges, indicating that manufacturing tolerant designs are achieved. Table \ref{First_set} summarizes the results; the maximum stress constraint violation is given for the points within the $\eta$ range considered during optimization, i.e., between the vertical dashed lines in the post-processing stress graphs, and for $\eta = 0.5$ for the deterministic case.

\begin{table}[ht!]
\centering
\caption{Volume fractions of intermediate topologies, maximum stress constraint violations, and number of iterations for the cases in Figure \ref{L1_topologies}. Run times obtained using $2$ nodes on the DTU Sophia cluster.}
\begin{tabular}{lclcccc}
\hline
Filter type & $N_{j}$ & $\eta \in \left\lbrace \eta_{min} , ... , \eta_{max} \right\rbrace$ & Vol. frac. & $\max \left( \frac{\sigma_{max}}{\sigma_{y}} - 1 \right)$ & Iterations & Run time (h) \\
\hline
Single & $1$ & $\eta = 0.5$ & $10.29 \%$ & $0.95 \%$ & $500$ & $7.1$ \\
Double & $3$ & $\eta \in \left\lbrace 0.35 , 0.5 , 0.65 \right\rbrace$ & $13.13 \%$ & $2.70 \%$ & $500$ & $16.3$ \\
Double & $3$ & $\eta \in \left\lbrace 0.2  , 0.5 , 0.8  \right\rbrace$ & $15.11 \%$ & $2.07 \%$ & $500$ & $15.6$ \\
\hline
\end{tabular}
\label{First_set}
\end{table}

\begin{figure}[ht!]
\centering
\includegraphics[width=0.85\textwidth]{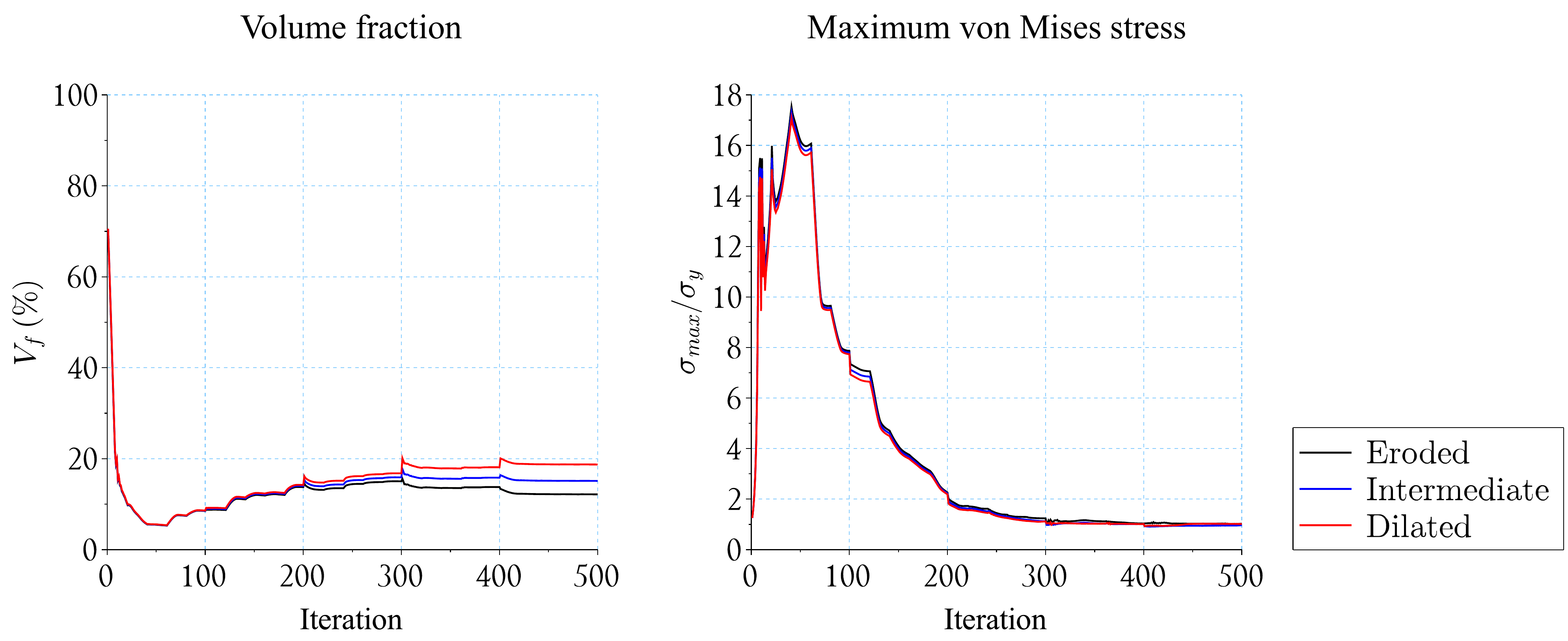}
\caption{Structural volume (left) and maximum von Mises stress (right) convergence histories for the results shown in Figure \ref{L1_DF_eta03_topologies}.}
\label{Historic_of_convergence}
\end{figure}

Figure \ref{Historic_of_convergence} shows the convergence history of both the volume fraction (left) and the maximum von Mises equivalent stress (right) for eroded, intermediate and dilated designs from Figure \ref{L1_DF_eta03_topologies}. Analyzing the volume convergence, one can verify strong volume minimization at the first few iterations; this is justified, in this case, since the optimization procedure is started with a very small penalization parameter, leading to very small weight on the stress constraints. As the Lagrange multipliers and penalization parameter are updated, the volume is increased, since more material is required to satisfy the stress constraints. Note that the differences between eroded and intermediate, and between intermediate and dilated structural volumes increase every $100$ iterations, after each $\beta$ update.

Analyzing the stress convergence, one can observe very large stress values at the first $200$ iterations; this is justified, since in the beginning of the iterative procedure the problem is volume dominated rather than stress. After a few updates of Lagrange multipliers and penalization parameter, the maximum stress values decrease, since the weight of the stress constraints on the augmented Lagrangian function is increased. One can verify smooth convergence of the maximum von Mises stress after $300$ iterations, for eroded, intermediate and dilated designs.

\subsection{Robust single filter \texorpdfstring{$\times$}{Lg} Robust double filter}\label{s411}

In this subsection, we show what happens if the standard single filter is used instead of the double filter in the robust formulation, i.e., $\overline{\bm{\rho}}$ instead of $\overline{\overline{\bm{\rho}}}$ in Equation \eqref{Problema_otm_3}. The same L-shaped design problem is addressed. In this example, we use $R = 0.08$, and hence $\beta_{max} = 7.390 \cong \beta_{lim}^{PDE}/2$. The iterative procedure is started with the first $\beta \leqslant 1$ when successively dividing $\beta_{max}$ by two, i.e., $\beta = \frac{7.390}{2^{3}}$.

\begin{figure}[ht!]
\centering
\includegraphics[width=0.99\textwidth]{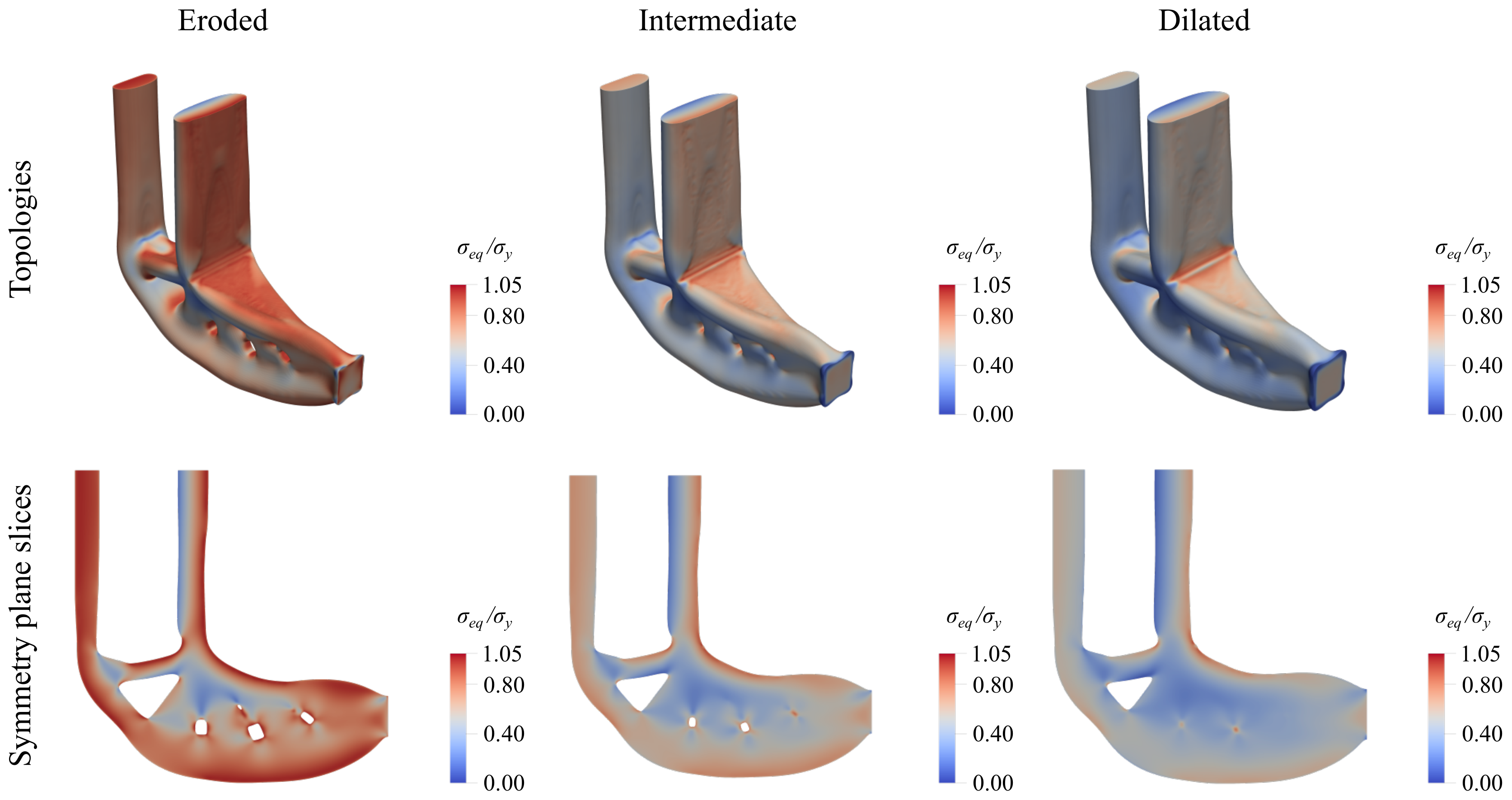}
\caption{Robust single filter, $\eta \in \left\lbrace 0.3 , 0.5 , 0.7 \right\rbrace$. Eroded, intermediate and dilated topologies (first row) and symmetry plane slices (second row).}
\label{L1_eta020_3pt}
\end{figure}

\begin{figure}[ht!]
\centering
\includegraphics[width=0.99\textwidth]{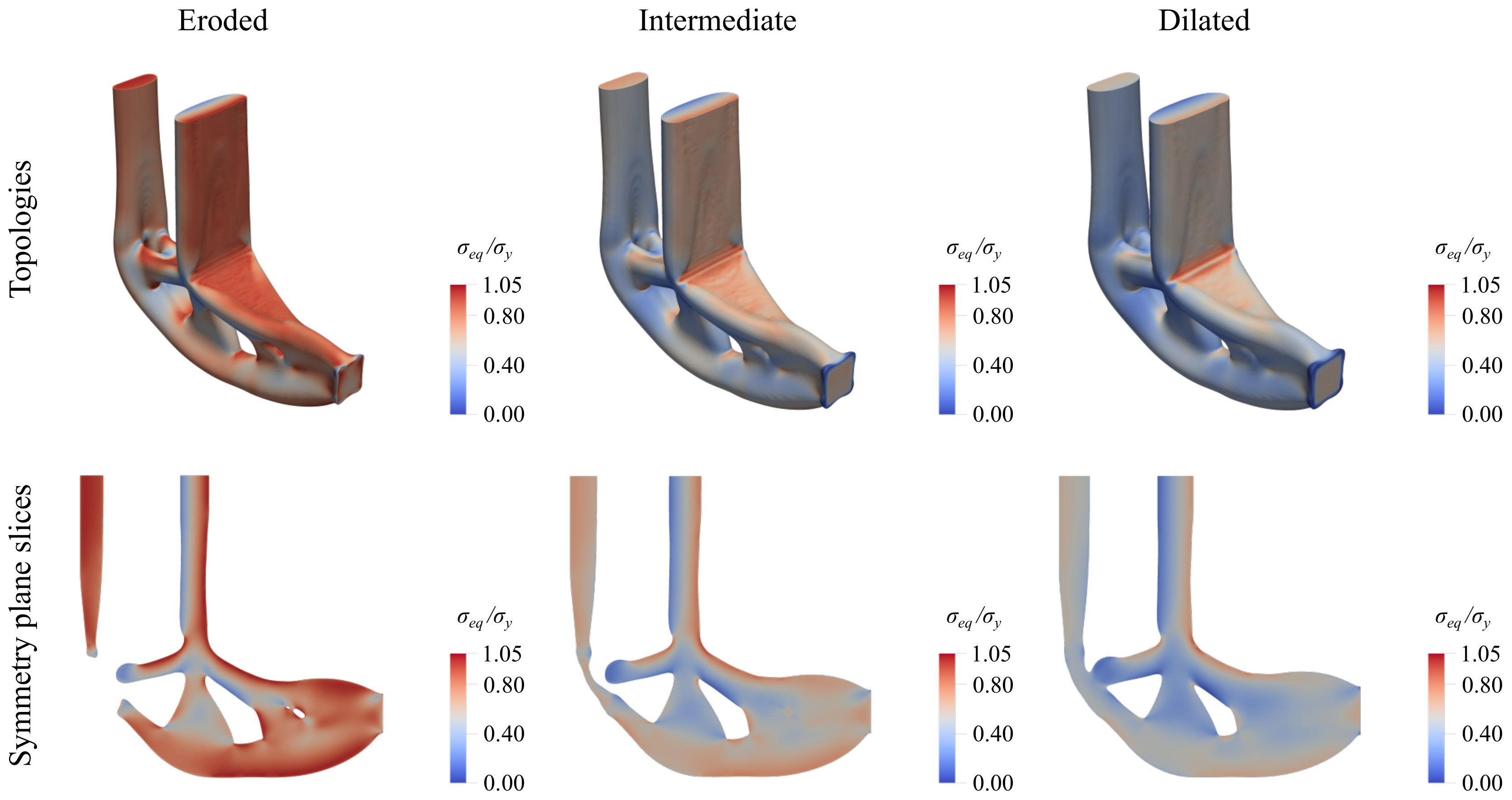}
\caption{Robust single filter, $\eta \in \left\lbrace 0.3 , 0.4 , 0.5 , 0.6 , 0.7 \right\rbrace$. Eroded, intermediate and dilated topologies (first row) and symmetry plane slices (second row).}
\label{L1_eta020_5pt}
\end{figure}

\begin{figure}[ht!]
\centering
\includegraphics[width=0.99\textwidth]{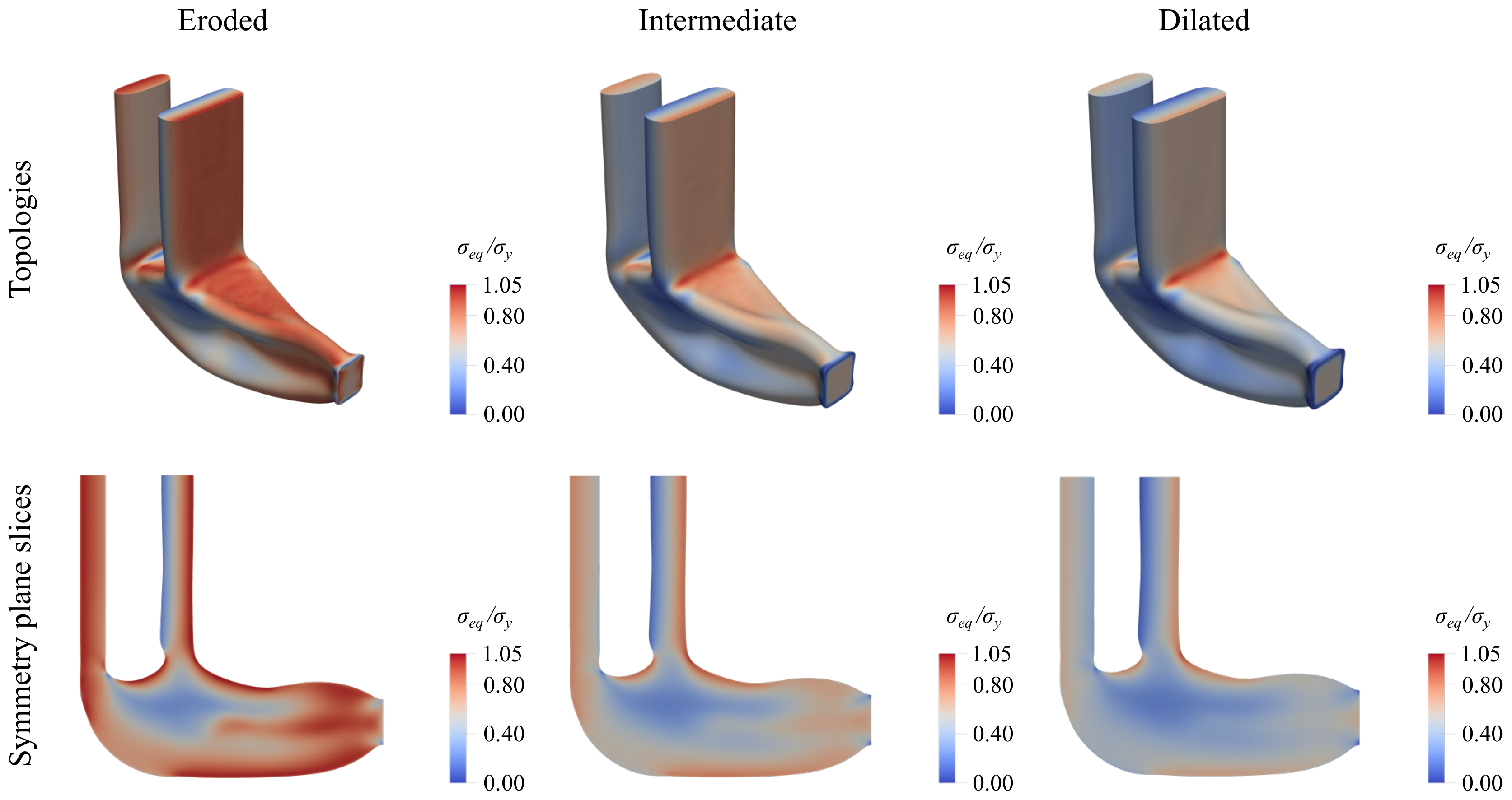}
\caption{Robust double filter, $\eta \in \left\lbrace 0.3 , 0.5 , 0.7 \right\rbrace$. Eroded, intermediate and dilated topologies (first row) and symmetry plane slices (second row).}
\label{L1_eta020_3pt_DF}
\end{figure}

\begin{figure}[ht!]
\centering
\includegraphics[width=0.99\textwidth]{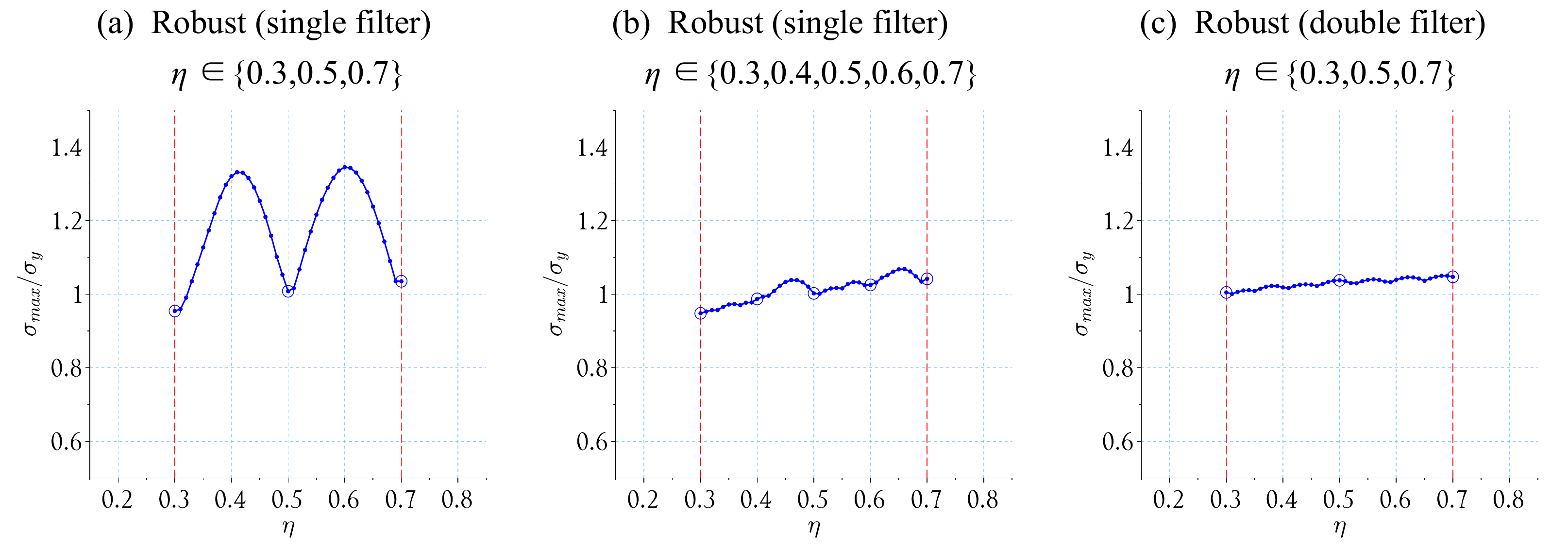}
\caption{(a), (b) and (c): post-processing stress graphs for topologies of Figures \ref{L1_eta020_3pt}, \ref{L1_eta020_5pt} and \ref{L1_eta020_3pt_DF}, respectively.}
\label{L1_stresses_R_008}
\end{figure}

Three optimization problems are solved: (1) robust single filter, for $N_{j} = 3$ and $\eta \in \left\lbrace 0.3 , 0.5 , 0.7 \right\rbrace$; (2) robust single filter, for $N_{j} = 5$ and $\eta \in \left\lbrace 0.3 , 0.4, 0.5 , 0.6 , 0.7 \right\rbrace$; and (3) robust double filter, for $N_{j} = 3$ and $\eta \in \left\lbrace 0.3 , 0.5 , 0.7 \right\rbrace$. Figures \ref{L1_eta020_3pt}, \ref{L1_eta020_5pt} and \ref{L1_eta020_3pt_DF} show the optimized topologies and respective symmetry plane slices, for cases (1), (2) and (3), respectively. Figure \ref{L1_stresses_R_008} shows the voxel-based post-processed stress graphs. When $N_{j} = 5$, eroded and dilated refer to the extremes of the employed $\eta$ range, i.e., the most eroded and most dilated designs.  Table \ref{Second_set} summarizes the results.

\begin{table}[ht!]
\centering
\caption{Volume fractions of intermediate topologies, maximum stress constraint violations, and number of iterations for the cases in Figures \ref{L1_eta020_3pt}, \ref{L1_eta020_5pt} and \ref{L1_eta020_3pt_DF}. Run times obtained using $2$ nodes on the DTU Sophia cluster.}
\begin{tabular}{lclcccc}
\hline
Filter type & $N_{j}$ & $\eta \in \left\lbrace \eta_{min} , ... , \eta_{max} \right\rbrace$ & Vol. frac. & $\max \left( \frac{\sigma_{max}}{\sigma_{y}} - 1 \right)$ & Iterations & Run time (h) \\
\hline
Single & $3$ & $\eta \in \left\lbrace 0.3  , 0.5 , 0.7  \right\rbrace$ & $16.55 \%$ & $34.53 \%$ & $400$ & $14.0$ \\
Single & $5$ & $\eta \in \left\lbrace 0.3 , 0.4 , 0.5 , 0.6 , 0.7 \right\rbrace$ & $16.49 \%$ & $6.83 \%$ & $400$ & $23.8$ \\
Double & $3$ & $\eta \in \left\lbrace 0.3  , 0.5 , 0.7  \right\rbrace$ & $19.99 \%$ & $5.00 \%$ & $400$ & $13.8$ \\
\hline
\end{tabular}
\label{Second_set}
\end{table}

Figure \ref{L1_eta020_3pt} shows an important weakness of the robust single filter approach: eroded, intermediate, and dilated topologies are not the same; there are small holes in both eroded and intermediate designs that are not present in the dilated topology. When different topologies are obtained, eroded and dilated designs do not represent uniform boundary variations. Analyzing Figure \ref{L1_stresses_R_008} (a), one can verify that stress constraints are satisfied only for eroded, intermediate and dilated topologies; stress constraint violations up to $34.53 \%$ are verified between the control points, indicating that the intermediate topology is not robust at all in this case.

Figure \ref{L1_eta020_5pt} shows the solution to the same problem, but now for five fields of relative densities instead of three. In this case, the topological differences are still present, but we get a much better stress graph, as illustrated in Figure \ref{L1_stresses_R_008} (b). It should be emphasized, however, that although a nice stress graph is obtained, the robustness of the optimized result is still questionable, due to the differences among the topologies that still exist. Even though the finite element mesh is very fine, the employed voxel-based model is not suitable to accurately quantify some topological changes that occur in this case, as the nucleation of holes.

Analyzing the results for the robust double filter approach, Figures \ref{L1_eta020_3pt_DF} and Figure \ref{L1_stresses_R_008} (c), we can verify identical topologies and a very smooth stress behavior within the whole $\eta$ range considered during optimization, even though only three fields of relative densities are considered. In this case, a truly manufacturing tolerant design is obtained, since eroded and dilated topologies represent uniform boundary variations, and the stress constraints are satisfied for the whole $\eta$ range. Moreover, the CPU cost is effectively $3/5$ of that with five realizations and single filter.

\subsection{Mesh dependence study}\label{s42}

\begin{figure}[ht!]
\centering
\includegraphics[width=0.99\textwidth]{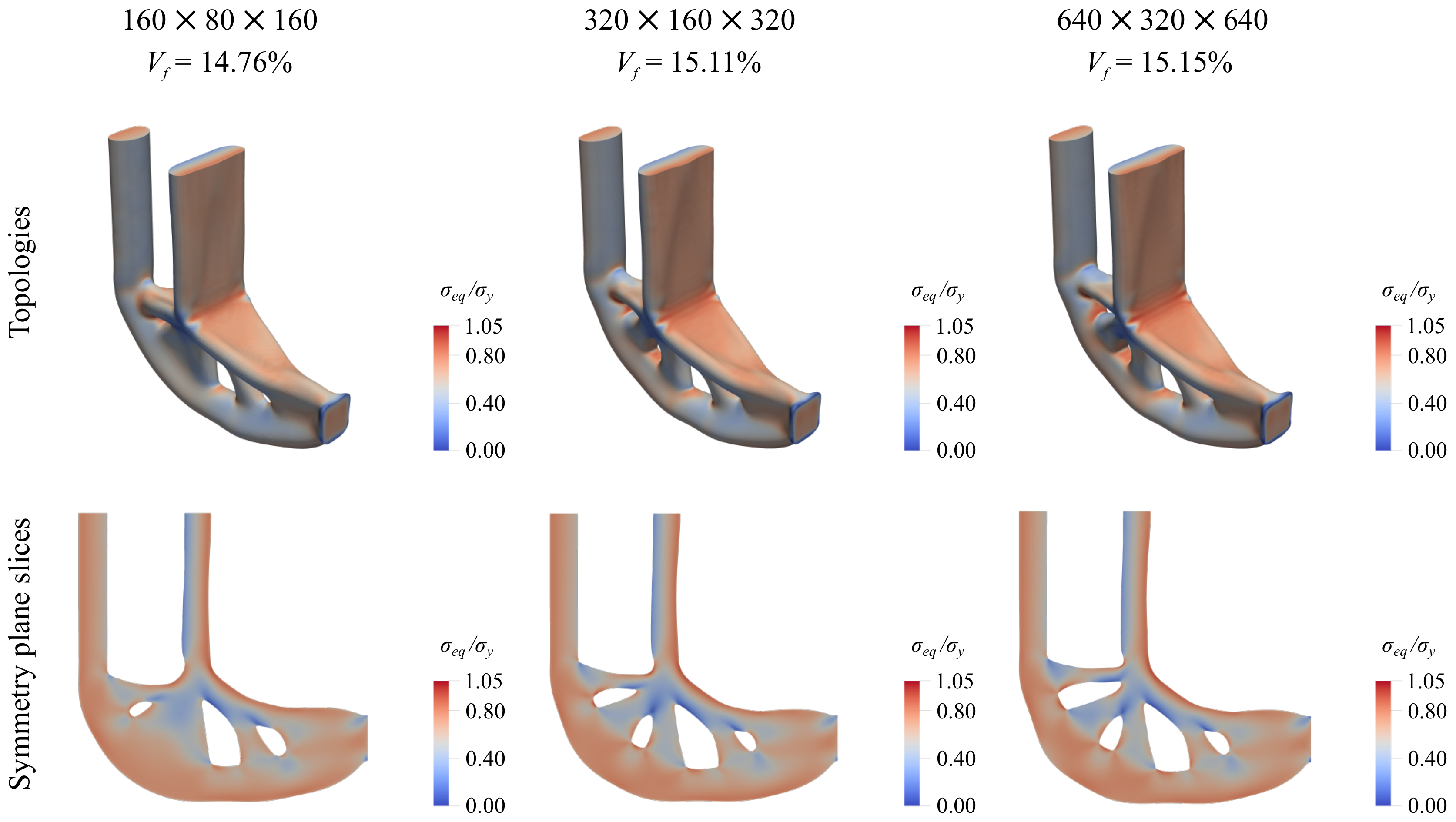}
\caption{Intermediate topologies (first row) and respective symmetry plane slices (second row), for different mesh sizes.}
\label{L1_mesh_dependence}
\end{figure}

\begin{figure}[ht!]
\centering
\includegraphics[width=0.99\textwidth]{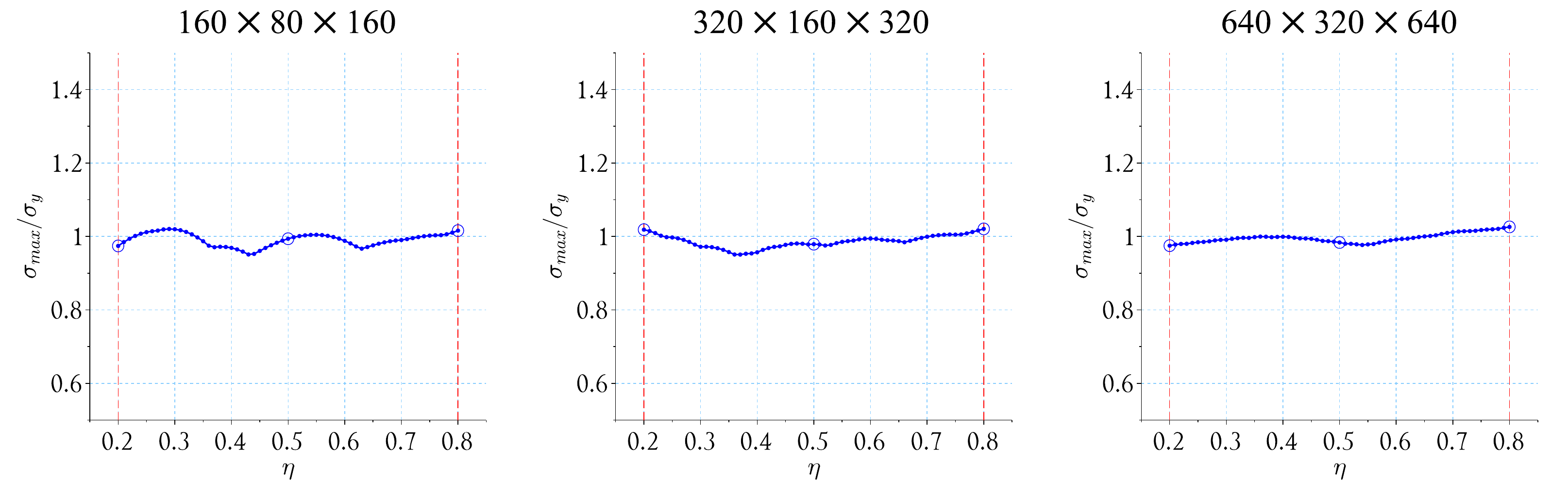}
\caption{Post-processing stress graphs for topologies of Figure \ref{L1_mesh_dependence}.}
\label{L1_mesh_dependence_stress}
\end{figure}

In this subsection, a mesh dependence study is performed. The robust double filter approach is used, with $R = 0.04$ and $\eta \in \left\lbrace 0.2 , 0.5 , 0.8 \right\rbrace$. Three meshes are employed: (1) $160 \times 80 \times 160 \cong 2.0$ million elements; (2) $320 \times 160 \times 320 \cong 16.4$ million elements; and (3) $640 \times 320 \times 640 \cong 131.1$ million elements. The value of $\beta_{max}$ is set based on the medium size mesh, as $\beta_{max} = 3.695 \cong \beta_{lim}^{PDE}/2$; this value corresponds to $\beta_{lim}^{PDE}$ and $\beta_{lim}^{PDE}/4$ for the coarser and finer mesh, respectively. The iterative procedure is started with $\beta = \frac{3.695}{2^{4}}$, and all topologies are obtained at $500$ iterations. Run times: $1.8$ h, $15.6$ h, and $11.9$ h; for $160 \times 80 \times 160$, $320 \times 160 \times 320$, and $640 \times 320 \times 640$ discretizations, respectively. These are solved using $2$, $2$ and $32$ nodes on the DTU Sophia cluster, resulting in $3.6$, $31.2$ and $380.8$ hour $\times$ node, respectively. Figure \ref{L1_mesh_dependence} shows the optimized intermediate topologies and respective slice views of the symmetry plane. Figure \ref{L1_mesh_dependence_stress} shows the respective voxel-based post-processing stress graphs.

Analyzing Figure \ref{L1_mesh_dependence}, one can verify identical topologies for medium and finer meshes, whereas a different topology is observed for the coarser mesh (one hole less). This is justified, since different mesh sizes provide different stress accuracy, which can lead the iterative procedure to slightly different topologies. When analyzing the volume fractions, one can verify a very nice agreement among all the results. Analyzing the stress graphs, illustrated in Figure \ref{L1_mesh_dependence_stress}, one can verify stress constraint feasibility for the whole $\eta$ range, for all mesh sizes. Maximum stress constraint violations are: $2.04 \%$, $2.07 \%$ and $2.61 \%$, for $160 \times 80 \times 160$, $320 \times 160 \times 320$ and $640 \times 320 \times 640$ discretizations, respectively. A very subtle non-smooth behavior is observed for the coarser mesh though, where $\beta_{max} \cong \beta_{lim}^{PDE}$ is employed. This behavior is justified, since the suggested value of $\beta_{max} \cong \beta_{lim}^{PDE}/2$ is slightly exceeded in this case \citep{Artigo6}, and hence the layer of intermediate material between solid and void phases is diminished, which in turn may lead to stress oscillation after uniform boundary variation.

\subsection{Study on the influence of filter radius}\label{s43}

This subsection aims at demonstrating the influence of the filter radius on the optimized topology. The L-shaped problem is addressed with the robust double filter approach. A mesh of $640 \times 320 \times 640$ is employed for all cases, with $N_{j} = 3$ and $\eta \in \left\lbrace 0.2 , 0.5 , 0.8 \right\rbrace$. Optimization is performed for three filtering radii: (1) $R = 0.02$; (2) $R = 0.04$; and (3) $R = 0.06$; with $\beta_{max} \cong \beta_{lim}^{PDE}/2$ in all cases. These are given by: (1) $\beta_{max} = 3.695$; (2) $\beta_{max} = 7.390$; and (3) $\beta_{max} = 11.085$. The iterative procedure is started with the first $\beta \leqslant 0.25$ when successively dividing $\beta_{max}$ by two.

\begin{figure}[ht!]
\centering
\includegraphics[width=0.99\textwidth]{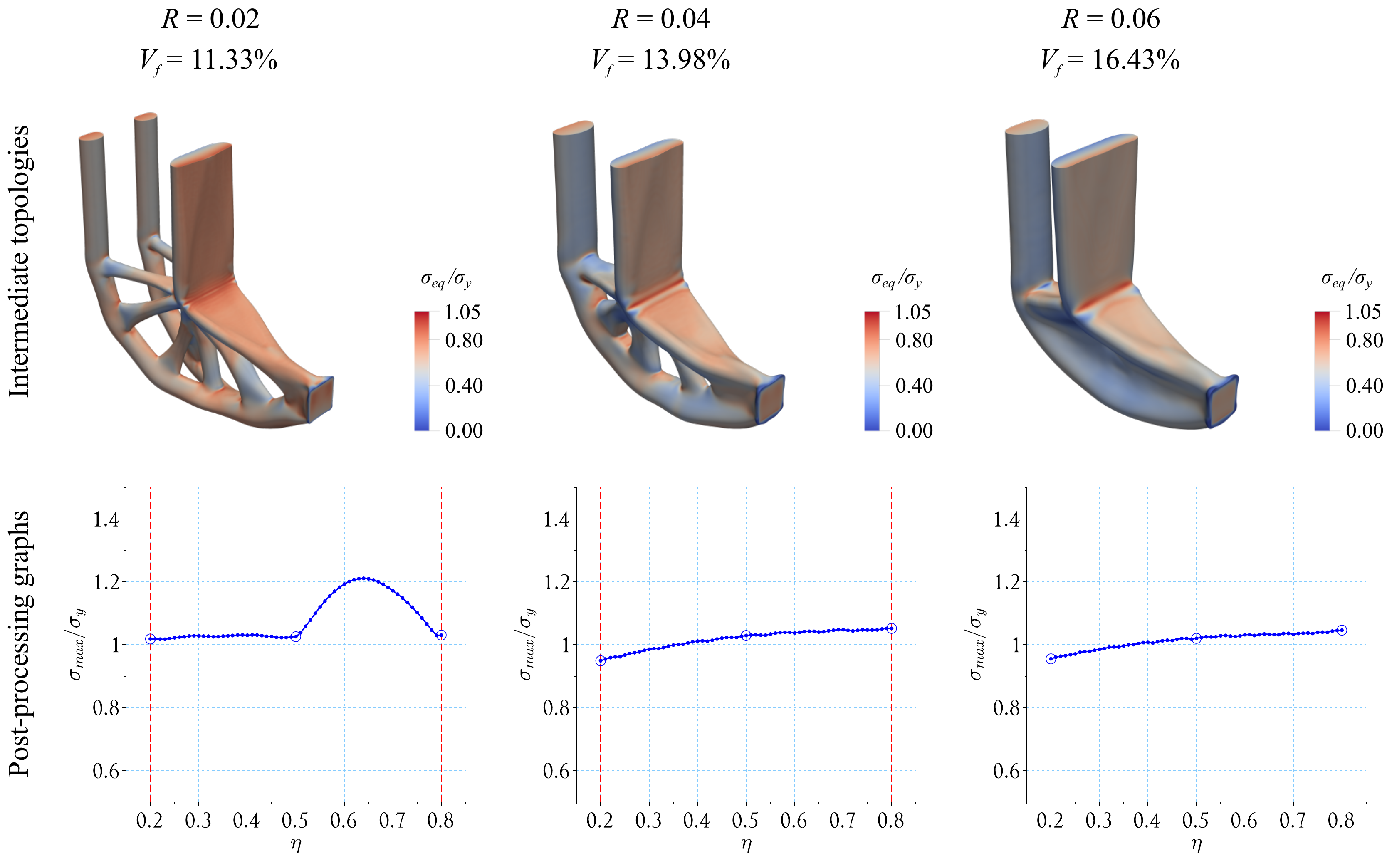}
\caption{Optimized intermediate topologies (first row) and respective post-processing stress graphs (second row) for different values of $R$.}
\label{L1_length_scale}
\end{figure}

Figure \ref{L1_length_scale} shows the optimized intermediate topologies and respective voxel-based post-processing stress graphs. Analyzing the topologies, one can verify that the larger the value of $R$, the smaller the number of structural details, and the larger the structural volume of the optimized intermediate topology. This is justified, since for fixed $\eta$ ranges, increasing the value of $R$ increases the minimum length scale on solid and void regions (Appendix \ref{a2}), making it more difficult to obtain small structural details. Besides, the larger the minimum length scale, the more restricted the space of admissible solutions becomes, making it more difficult to obtain a structure with low volume.

\begin{figure}[ht!]
\centering
\includegraphics[width=0.65\textwidth]{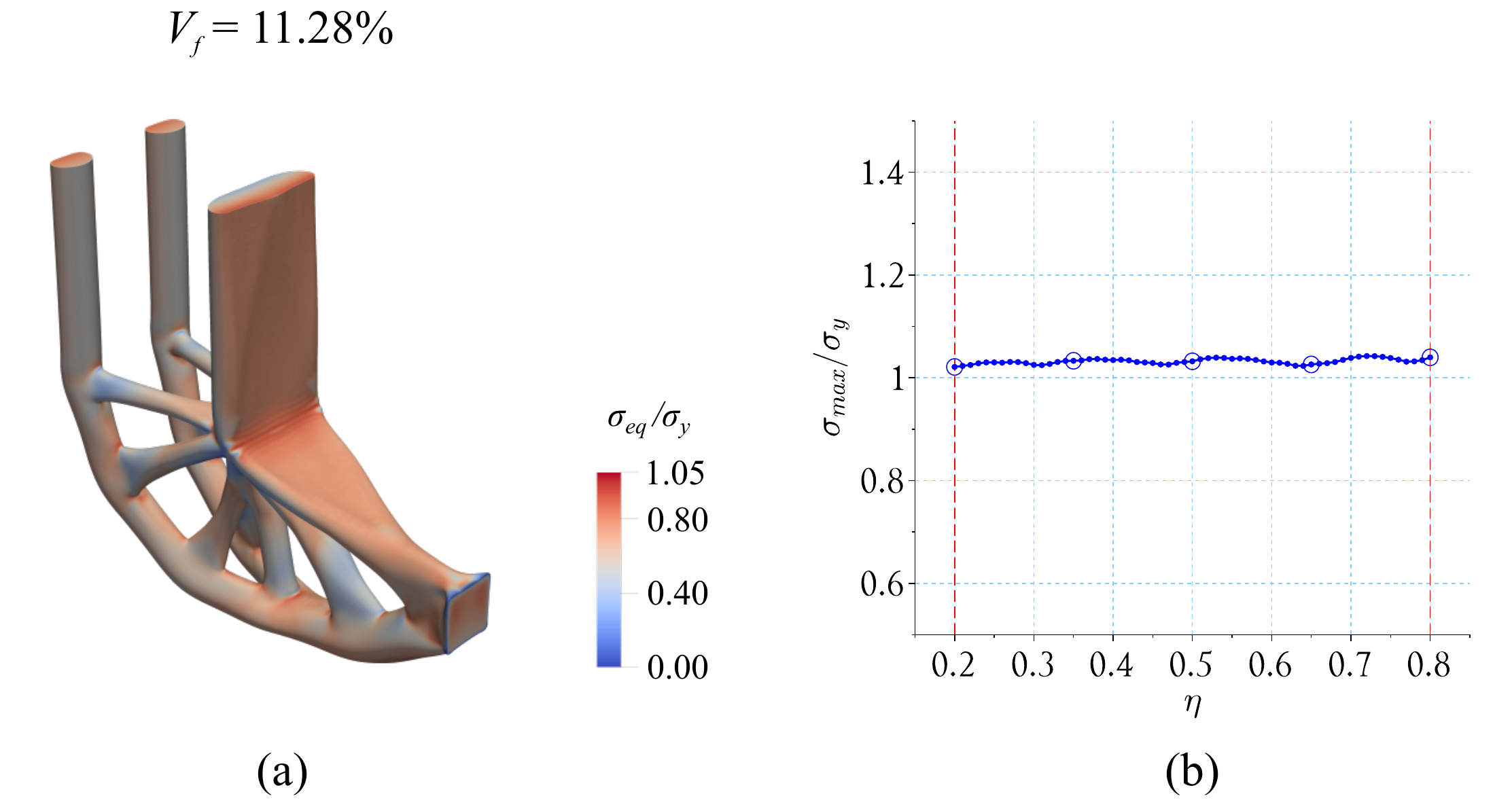}
\caption{(a) Optimized intermediate (robust) topology and (b) respective post-processing stress graph. Robust double filter for $\eta \in \left\lbrace 0.2 , 0.35 , 0.5 , 0.65 , 0.8 \right\rbrace$.}
\label{L1_eta030_5pt_DF_R002}
\end{figure}

Analyzing the stress graphs in Figure \ref{L1_length_scale}, one observes very smooth stress behavior for $R = 0.04$ and $R = 0.06$, whereas a stress peak is observed between intermediate and eroded topologies for $R = 0.02$. The stress peak behavior occurs for the smallest filtering radius, and hence, the smallest minimum manufacturing tolerance. This behavior may seem counterintuitive at first, since smaller manufacturing tolerances seem easier to achieve; however, it is not totally unexpected. A similar behavior is observed by Da Silva et al. \cite{Artigo7}, when addressing two compliant mechanism problems with same mesh and different filter radii; in that case, a stress peak was observed between dilated and intermediate designs for the case with smaller $R$, whereas an absolutely smooth behavior was observed for the case with larger $R$. Observation of these particular cases indicates that topologies with more structural details and thinner structural members are more sensitive to boundary variations when stress requirements are taken into account in the formulation. In order to handle this issue, in this paper, the problem is solved again for $N_{j} = 5$ and $\eta \in \left\lbrace 0.2 , 0.35 , 0.5 , 0.65 , 0.8 \right\rbrace$; topology and voxel-based stress graph are shown in Figure \ref{L1_eta030_5pt_DF_R002}.


Comparing the topologies for $N_{j} = 3$ and $N_{j} = 5$, given by Figures \ref{L1_length_scale} (for $R = 0.02$) and \ref{L1_eta030_5pt_DF_R002} (a), respectively, one can observe no significant difference; the topologies are the same, with only a few subtle differences on their shapes. Analyzing Figure \ref{L1_eta030_5pt_DF_R002} (b), however, one can verify that the stress peak behavior was totally suppressed, and now an absolutely smooth stress behavior is observed, indicating that this problem is highly sensitive to boundary variation. It is unfortunate, however, that the number of necessary fields of physical relative densities is not known in advance, and that it is not possible to provide a general rule estimating the number of fields required for all cases based on the results obtained in this single example. It should be noted, however, that this was the only case in which more than three realizations were necessary to obtain an absolutely smooth stress behavior after uniform boundary variations when using the double filter approach. All the other results throughout the manuscript were obtained with three fields only.

Table \ref{tabela_escala_comprimento} summarizes the results obtained in this subsection. The difference among the number of iterations comes from the different $\beta_{max}$ values considered. Since all iterative procedures started with similar $\beta$ values, those with a higher upper bound take more iterations to reach the convergence criteria.

\begin{table}[ht!]
\centering
\caption{Volume fractions of intermediate topologies, maximum stress constraint violations, and number of iterations for the cases in Figures \ref{L1_length_scale} and \ref{L1_eta030_5pt_DF_R002}. Run times obtained using $32$ nodes on the DTU Sophia cluster.}
\begin{tabular}{lcclcccc}
\hline
Filter type & $N_{j}$ & $R$ & $\eta \in \left\lbrace \eta_{min} , ... , \eta_{max} \right\rbrace$ & Vol. frac. & $\max \left( \frac{\sigma_{max}}{\sigma_{y}} - 1 \right)$ & Iterations & Run time (h) \\
\hline
Double & $3$ & $0.02$ & $\eta \in \left\lbrace 0.2  , 0.5 , 0.8  \right\rbrace$ & $11.33 \%$ & $21.09 \%$ & $500$ & $16.8$ \\
Double & $3$ & $0.04$ & $\eta \in \left\lbrace 0.2  , 0.5 , 0.8  \right\rbrace$ & $13.98 \%$ & $5.20 \%$ & $619$ & $13.8$ \\
Double & $3$ & $0.06$ & $\eta \in \left\lbrace 0.2  , 0.5 , 0.8  \right\rbrace$ & $16.43 \%$ & $4.63 \%$ & $700$ & $24.2$ \\
Double & $5$ & $0.02$ & $\eta \in \left\lbrace 0.2  , 0.35 , 0.5 , 0.65 , 0.8  \right\rbrace$ & $11.28 \%$ & $4.22 \%$ & $500$ & $28.9$ \\
\hline
\end{tabular}
\label{tabela_escala_comprimento}
\end{table}

\begin{figure}[ht!]
\centering
\includegraphics[width=0.99\textwidth]{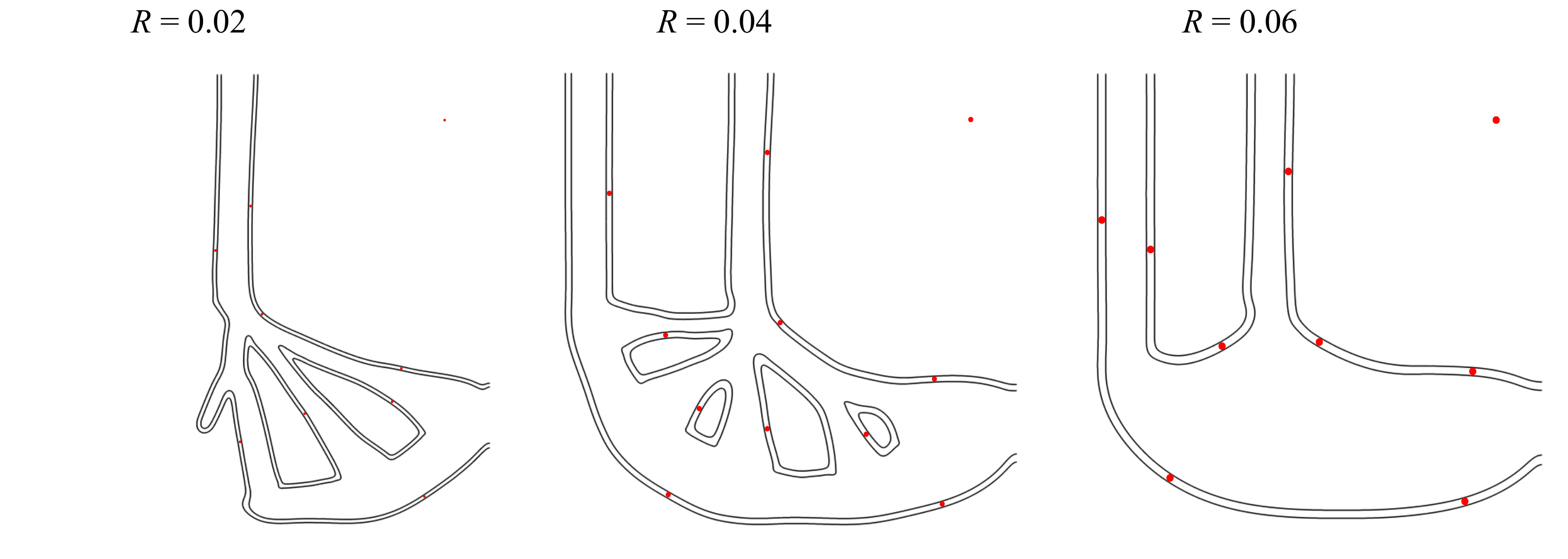}
\caption{Tolerance ranges. Eroded and dilated contour plots of symmetry plane slices for different $R$. Circles indicate the minimum expected distance between eroded and dilated designs. A few circles are included between eroded and dilated contour plots for better interpretation.}
\label{Topologies_length_scale}
\end{figure}

Figure \ref{Topologies_length_scale} shows eroded and dilated contour plots of symmetry plane slices together with the expected tolerance ranges, computed with the equations provided in Appendix \ref{a2}. The circles represent twice the minimum manufacturing tolerance, i.e., the minimum distance between eroded and dilated designs. Note that the bi-dimensional representation is adopted for better understanding and visualization. Actually, spherical manufacturing tolerances are obtained in this case, since three-dimensional problems are addressed \citep{Wang2011}. Analyzing Figure \ref{Topologies_length_scale}, one can verify an excellent agreement between expected and obtained tolerance ranges.


\subsection{Body-fitted post-processing}\label{s44}

\begin{figure}[ht!]
\centering
\includegraphics[width=0.99\textwidth]{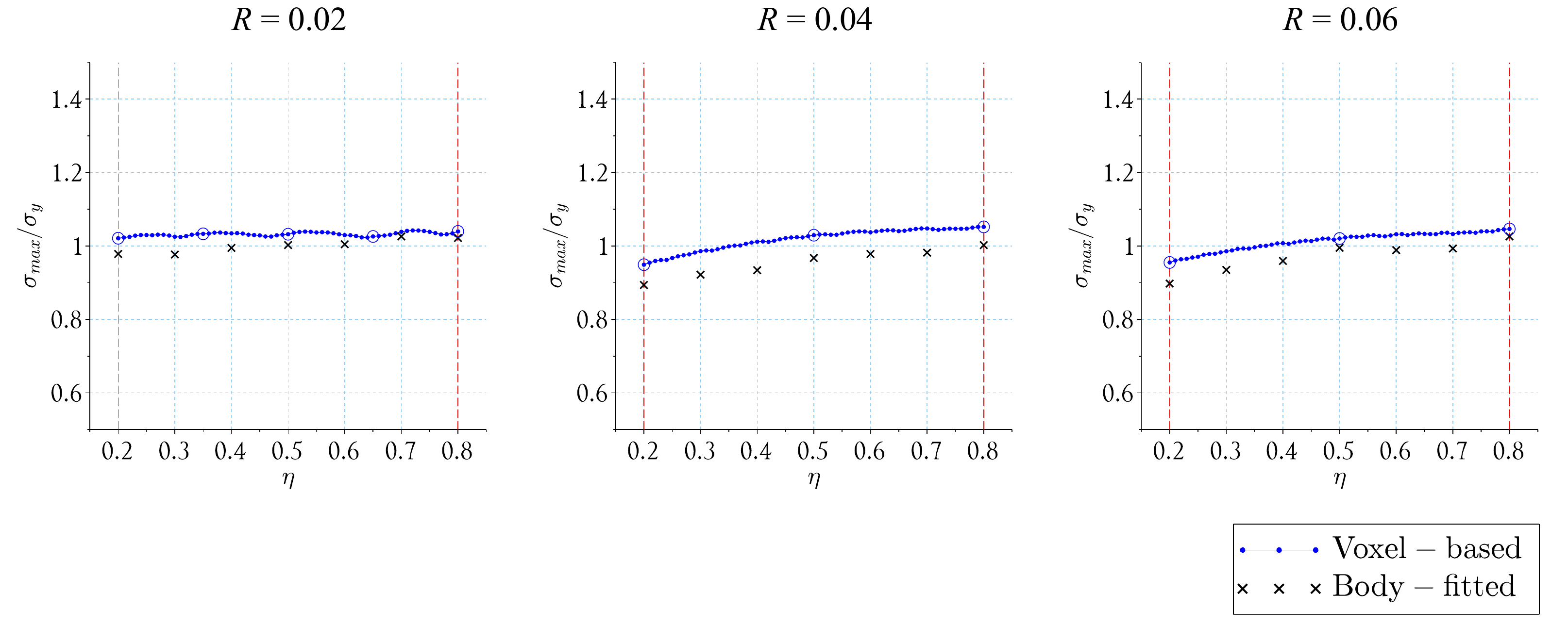}
\caption{Voxel-based and body-fitted stress graphs for $R = 0.02$ (left), $R = 0.04$ (middle), and $R = 0.06$ (right).}
\label{Stress_compara_COMSOL}
\end{figure}

In order to verify the accuracy of the voxel-based stress response, we perform a body-fitted post-processing scheme on some results from subsection \ref{s43} using the simulation software COMSOL Multiphysics 5.5. The smoothed designs are imported into COMSOL, and these are discretized with body-fitted meshes comprised of linear tetrahedral elements. The smoothed designs are extracted from the second level filtered densities using ParaView, for $\eta \in \left\lbrace 0.2 , 0.3 , ... , 0.7 , 0.8 \right\rbrace$. Figure \ref{Stress_compara_COMSOL} shows the post-processed body-fitted stress graphs together with the voxel-based graphs presented earlier in Figures \ref{L1_length_scale} (for $R = 0.04$ and $R = 0.06$) and \ref{L1_eta030_5pt_DF_R002} (for $R = 0.02$). The number of elements in each body-fitted mesh varies; the average number is approximately $890,000$.

Analyzing Figure \ref{Stress_compara_COMSOL}, one can verify an excellent agreement between voxel-based and body-fitted maximum von Mises stress values. Table \ref{Tabela_compara_COMSOL} shows maximum errors between both models and maximum stress constraint violations for the body-fitted meshes. The results indicate that: (1) the voxel-based model employed herein has good accuracy in stress computation; and (2) the proposed formulation is able to achieve truly manufacturing tolerant solutions which satisfy the stress failure criterion.

\begin{table}[ht!]
\centering
\caption{Maximum von Mises stress differences between voxel-based and body-fitted models, and maximum stress constraint violations for the body-fitted models, for $R = 0.02$, $R = 0.04$ and $R = 0.06$.}
\begin{tabular}{lclccc}
\hline
Filter type & $N_{j}$ & $\eta \in \left\lbrace \eta_{min}, ... , \eta_{max} \right\rbrace$ & $R$ & $\text{max} \left( \frac{\vert \bm{\sigma}^{pixel} - \bm{\sigma}^{fitted} \vert}{\bm{\sigma}^{fitted}} \right)$ & $\frac{\sigma_{max}^{fitted}}{\sigma_{y}}-1$ \\
\hline
Double & $5$ & $\eta \in \left\lbrace 0.2 , 0.35 , 0.5 , 0.65 , 0.8 \right\rbrace$ & $0.02$ & $4.95 \%$ & $2.60 \%$ \\
Double & $3$ & $\eta \in \left\lbrace 0.2 , 0.5 , 0.8 \right\rbrace$               & $0.04$ & $8.30 \%$ & $0.24 \%$ \\
Double & $3$ & $\eta \in \left\lbrace 0.2 , 0.5 , 0.8 \right\rbrace$               & $0.06$ & $6.41 \%$ & $2.60 \%$ \\
\hline
\end{tabular}
\label{Tabela_compara_COMSOL}
\end{table}

\begin{figure}[ht!]
\centering
\includegraphics[width=0.99\textwidth]{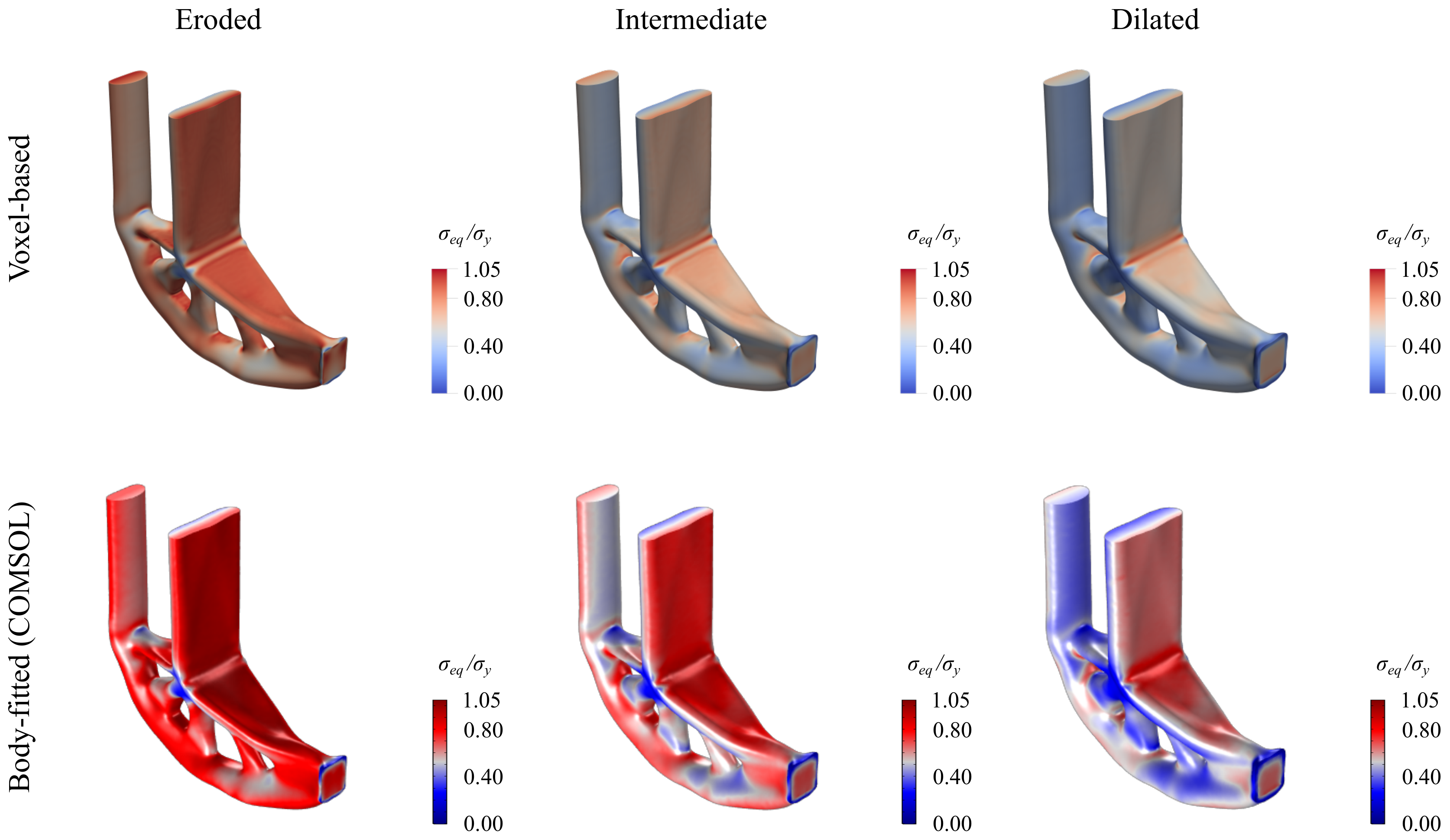}
\caption{L-shaped design problem for $R = 0.04$. Eroded, intermediate and dilated voxel-based and body-fitted (COMSOL) models with respective von Mises stress distributions.}
\label{L1_COMSOL_R004}
\end{figure}

\begin{figure}[ht!]
\centering
\includegraphics[width=0.67\textwidth]{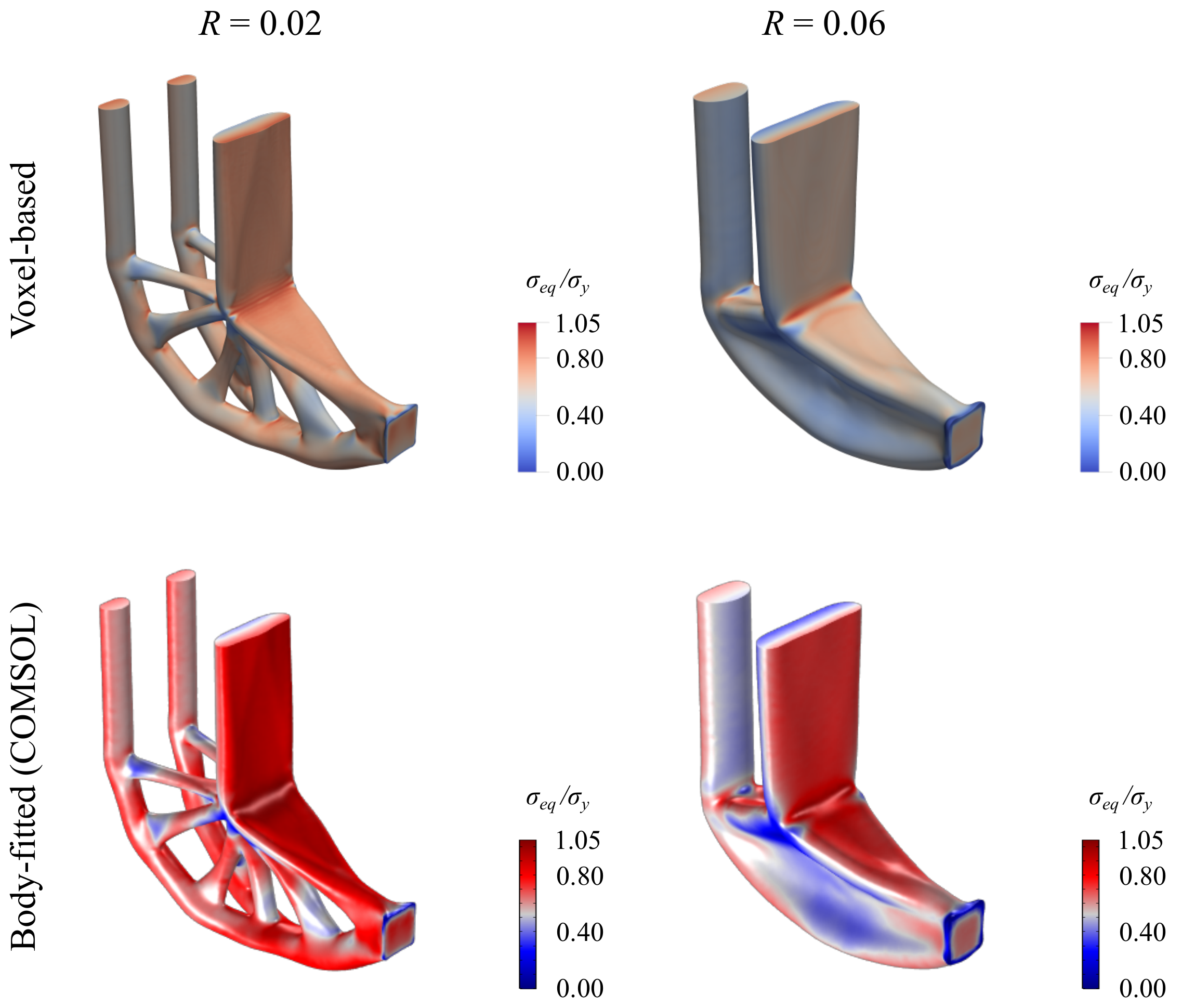}
\caption{L-shaped design problem for $R = 0.02$ (left) and $R = 0.06$ (right). Intermediate voxel-based and body-fitted (COMSOL) models with respective von Mises stress distributions.}
\label{L1_COMSOL_R002_R006}
\end{figure}

Figure \ref{L1_COMSOL_R004} shows voxel-based (first row) and body-fitted (second row) von Mises stress distributions for eroded, intermediate and dilated structures, for the case with $R = 0.04$. Although the color scales are slightly different from each other, one can observe a large extent of similarity between both stress models. First, when comparing each topology separately, e.g., the dilated one, one can observe a clear agreement in the way the stresses are distributed; the same applies to intermediate and eroded topologies. Second, the same tendency of having the eroded topology as the most stressed one and the dilated topology with fewer highly stressed regions is also observed in the body-fitted model. Figure \ref{L1_COMSOL_R002_R006} shows the intermediate topologies for the cases with $R = 0.02$ and $R = 0.06$. One can verify a clear agreement between voxel-based and body-fitted stress distributions also in these cases.

\subsection{Additional numerical results}\label{s45}

For the sake of completeness, three additional numerical examples are solved and further post-processed with body-fitted COMSOL models: (1) the L-shaped problem with very small length scale; (2) the L-shaped problem with different load orientation; and (3) a crack problem. The input data are defined in the beginning of section \ref{s4}, and they are the same for all problems, with the exception of $R$, $\eta$, mesh size, and boundary conditions; these are defined for each problem. All problems are addressed with the robust double filter approach and $N_{j} = 3$, i.e., three fields of relative densities. In all cases, the iterative procedure is started with the first $\beta \leqslant 0.25$ when successively dividing $\beta_{max}$ by two.

\subsubsection{L-shaped problem with very small length scale}

\begin{figure}[ht!]
\centering
\includegraphics[width=0.99\textwidth]{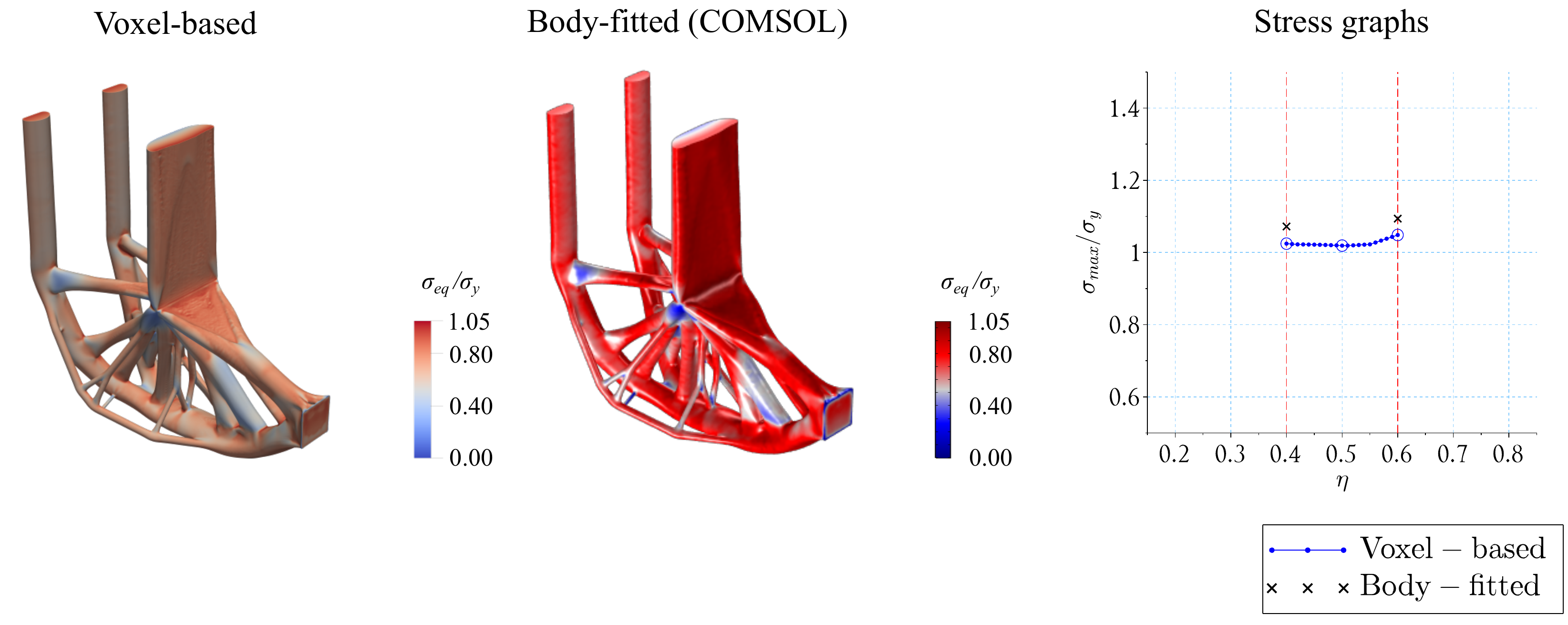}
\caption{Dilated topologies with respective von Mises stress distributions: voxel-based (left), body-fitted (middle); and post-processing stress graphs (right). L-shaped design problem for $\eta \in \left\lbrace 0.4 , 0.5 , 0.6 \right\rbrace$ and $R = 0.01$.}
\label{L1_COMSOL_eta010}
\end{figure}

In this subsection, the L-shaped design problem is solved for a very small length scale, in order to promote the appearance of structural details. Data: $R = 0.01$, $\eta \in \left\lbrace 0.4 , 0.5 , 0.6 \right\rbrace$ and $640 \times 320 \times 640$ (mesh size); with $\beta_{max} = 1.848 \cong \beta_{lim}^{PDE}/2$. The optimized design is obtained at $400$ iterations. Fine body-fitted meshes are employed for post-processing; the average number of elements is $3,560,000$. In this case, we were not able to perform the body-fitted verification for the intermediate design. It should be noted that meshing such complicated geometries with body-fitted models is notoriously difficult and sometimes not possible. Figure \ref{L1_COMSOL_eta010} shows the dilated topology (voxel-based and body-fitted stress models) and the post-processing stress graphs. Maximum stress constraint violations for voxel-based and body-fitted models are $4.85 \%$ and $9.38 \%$, respectively. Maximum error between both models is $4.43 \%$.

A very small tolerance range is ensured due to the very small filter radius and $\eta$ range employed. Only three realizations were necessary to provide very smooth stress behavior, contrary to what happened for $R = 0.02$ and $\eta \in \left\lbrace 0.2 , 0.5 , 0.8 \right\rbrace$, Figure \ref{L1_length_scale}, indicating that the necessity of more realizations is not associated with the length scale itself, but to a combination of filter size and $\eta$ range.

\subsubsection{L-shaped problem with different load orientation}

\begin{figure}[ht!]
\centering
\includegraphics[width=0.85\textwidth]{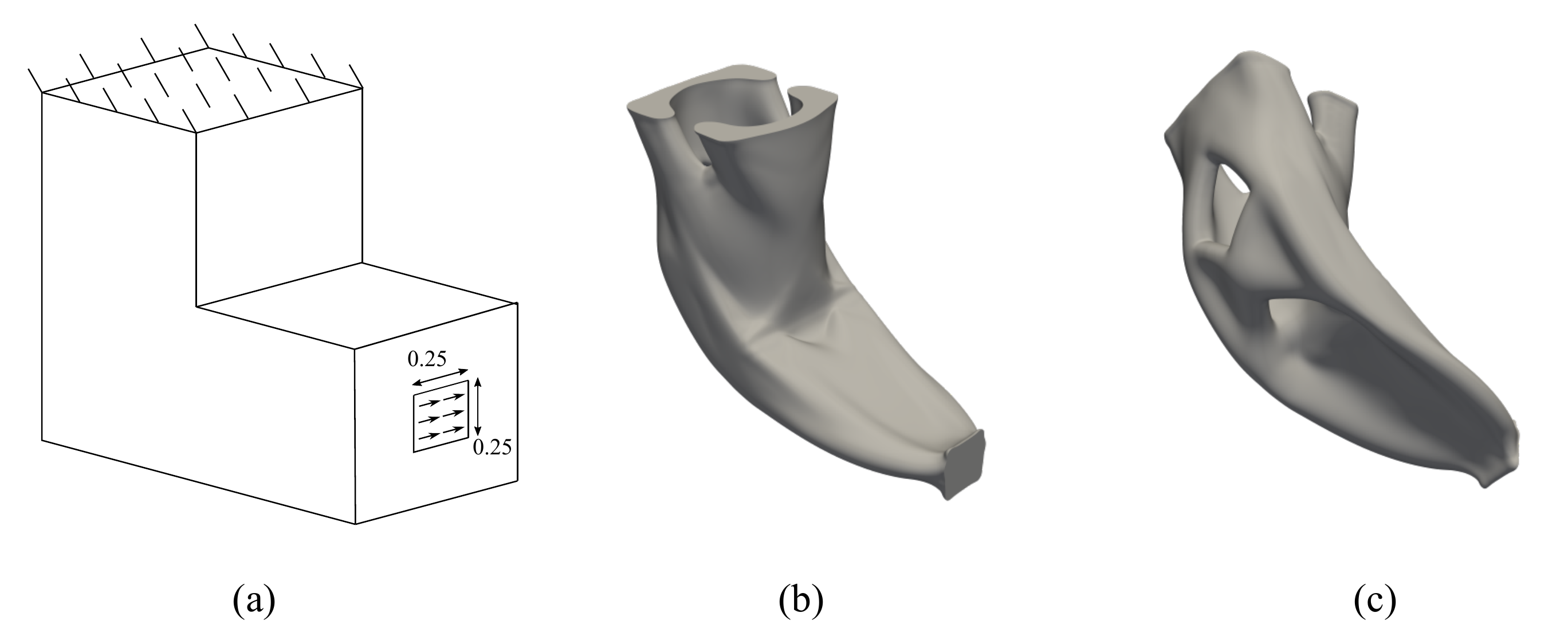}
\caption{(a) L-shaped design problem for different load orientation and support boundary condition; (b) and (c) robust (intermediate) optimized topology from two different points of view.}
\label{L4_problem_and_result}
\end{figure}

We have the same design domain as in Figure \ref{L1} (a), but different boundary conditions: (1) the roller condition is replaced by a fixed support condition; and (2) the applied load has a different direction, but same unitary magnitude. The non-design and internal padding regions are the same. The problem is solved for $R = 0.04$, $\eta \in \left\lbrace 0.2 , 0.5 , 0.8 \right\rbrace$ and a mesh of $640 \times 320 \times 640$ elements. In this case, $\beta_{max} = 7.390 \cong \beta_{lim}^{PDE}/2$. The optimized design is obtained at $600$ iterations and is illustrated in Figure \ref{L4_problem_and_result} (b) and (c), from two different points of view.

\begin{figure}[ht!]
\centering
\includegraphics[width=0.99\textwidth]{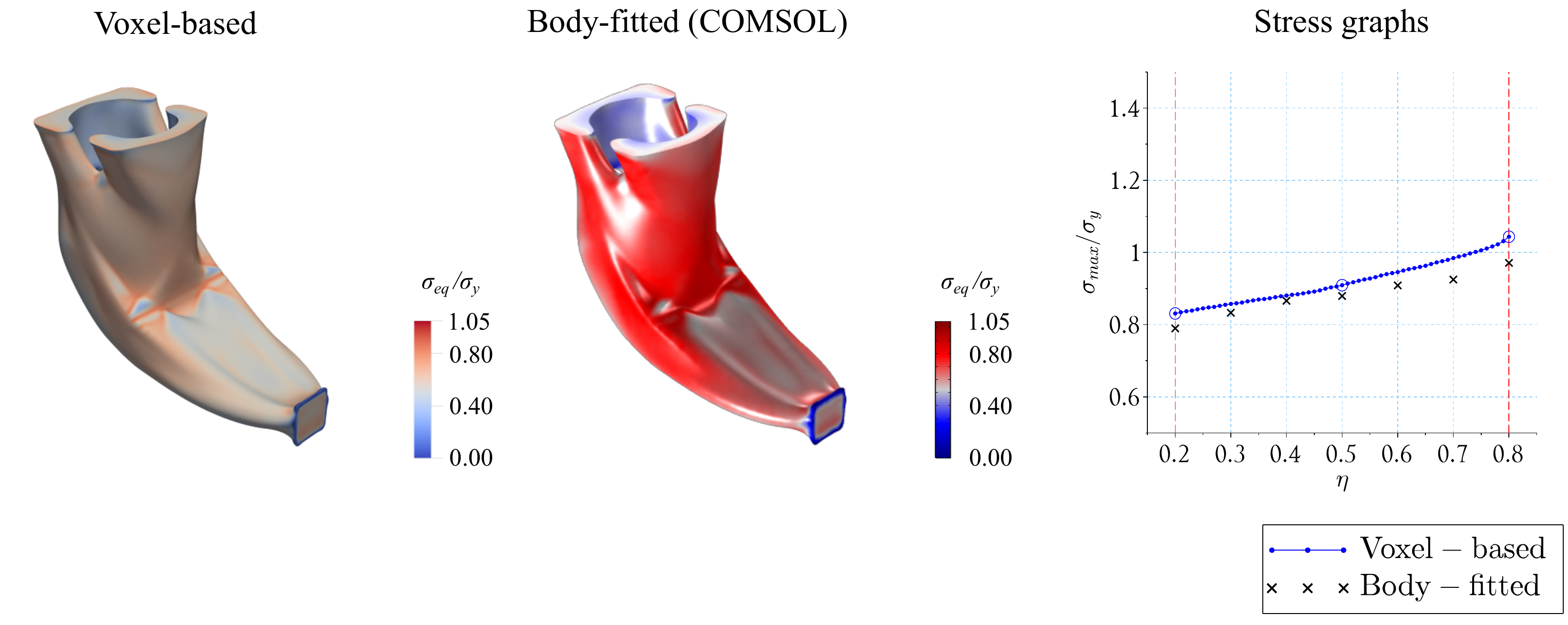}
\caption{Intermediate topologies with respective von Mises stress distributions: voxel-based (left), body-fitted (middle); and post-processing stress graphs (right). L-shaped design problem from Figure \ref{L4_problem_and_result}.}
\label{L4_COMSOL_R004}
\end{figure}

One observes an excellent agreement between voxel-based and body-fitted von Mises stress models, Figure \ref{L4_COMSOL_R004}. The average number of elements in the body-fitted meshes is $1,030,000$. Maximum stress constraint violation in the voxel-based model is $4.36 \%$ and maximum error between both models is $7.46 \%$. There is no stress constraint violation in the body-fitted model.


\subsubsection{Crack problem}

\begin{figure}[ht!]
\centering
\includegraphics[width=0.7\textwidth]{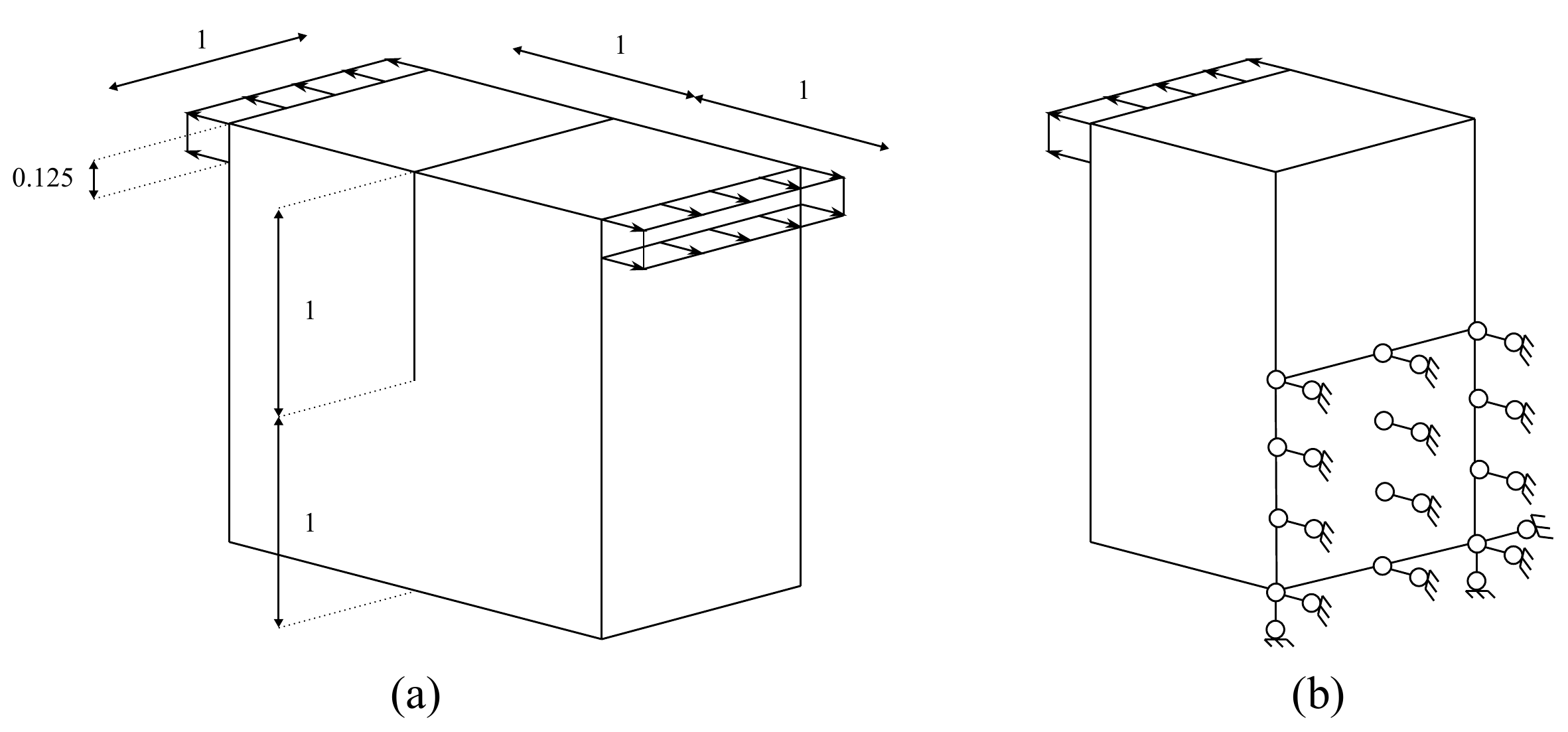}
\caption{(a) Crack problem; (b) Design domain employed in optimization (symmetry conditions).}
\label{C1}
\end{figure}

The discontinuous domain problem considered in this subsection is a three-dimensional version of the problem by Emmendoerfer and Fancello \cite{Fancello_Level_set2}. Figure \ref{C1} shows: (a) the original full problem; and (b) the reduced problem using symmetry addressed for the optimization. Applied load of unitary magnitude is distributed over a region of dimensions $1 \times 0.125 $. Internal padding regions are applied to surround the boundaries of the design domain, with the exception of regions near boundary conditions. The problem is solved for $R = 0.04$, $\eta \in \left\lbrace 0.2 , 0.5 , 0.8 \right\rbrace$ and a mesh of $400 \times 400 \times 800 = 128$ million elements, which corresponds to $256$ million for the full structure. In this case, $\beta_{max} = 9.238 \cong \beta_{lim}^{PDE}/2$. The optimized design is obtained at $702$ iterations.

\begin{figure}[ht!]
\centering
\includegraphics[width=0.99\textwidth]{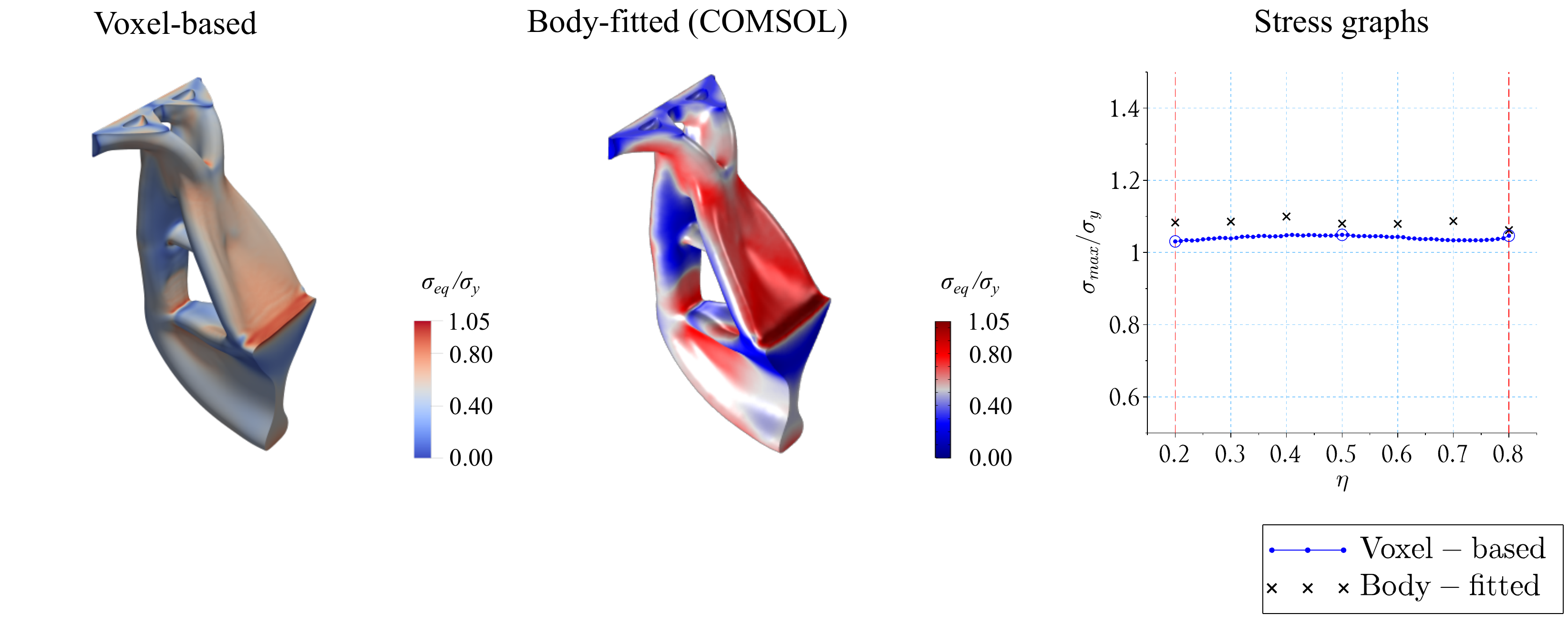}
\caption{Intermediate topologies with respective von Mises stress distributions: voxel-based (left), body-fitted (middle); and post-processing stress graphs (right). Crack problem (half model).}
\label{C1_COMSOL_R004}
\end{figure}

Figure \ref{C1_COMSOL_R004} shows the intermediate topology and the stress graphs; an excellent agreement between voxel-based and body-fitted models is observed. The average number of elements in the body-fitted meshes is $650,000$. Maximum stress constraint violations for voxel-based and body-fitted models are $4.89 \%$ and $9.96 \%$, respectively. Maximum error between both models is $4.91 \%$.


For the sake of completeness, the exact same structural problem is solved with the standard compliance-based robust formulation \citep{Wang2011,Clausen2017}, using the volume fraction obtained for the stress-constrained case, of $V_{f} = 15.23 \%$. The structural compliance of the eroded topology is minimized subject to a volume constraint on the dilated design. The prescribed volume fraction is updated every $20$ iterations, as $\overline{V}^{(d)} = \frac{\overline{V}^{(i)}}{V_{f}^{(i)}} V_{f}^{(d)}$, so that the volume constraint is satisfied for the intermediate design at the end of the optimization procedure \citep{Wang2011}. The resulting problem is solved with the Method of Moving Asymptotes \citep{MMA,Niels_MMA}, using the same $\beta$-continuation scheme as the stress-constrained case. The optimized design is obtained after $700$ iterations.

\begin{figure}[ht!]
\centering
\includegraphics[width=0.99\textwidth]{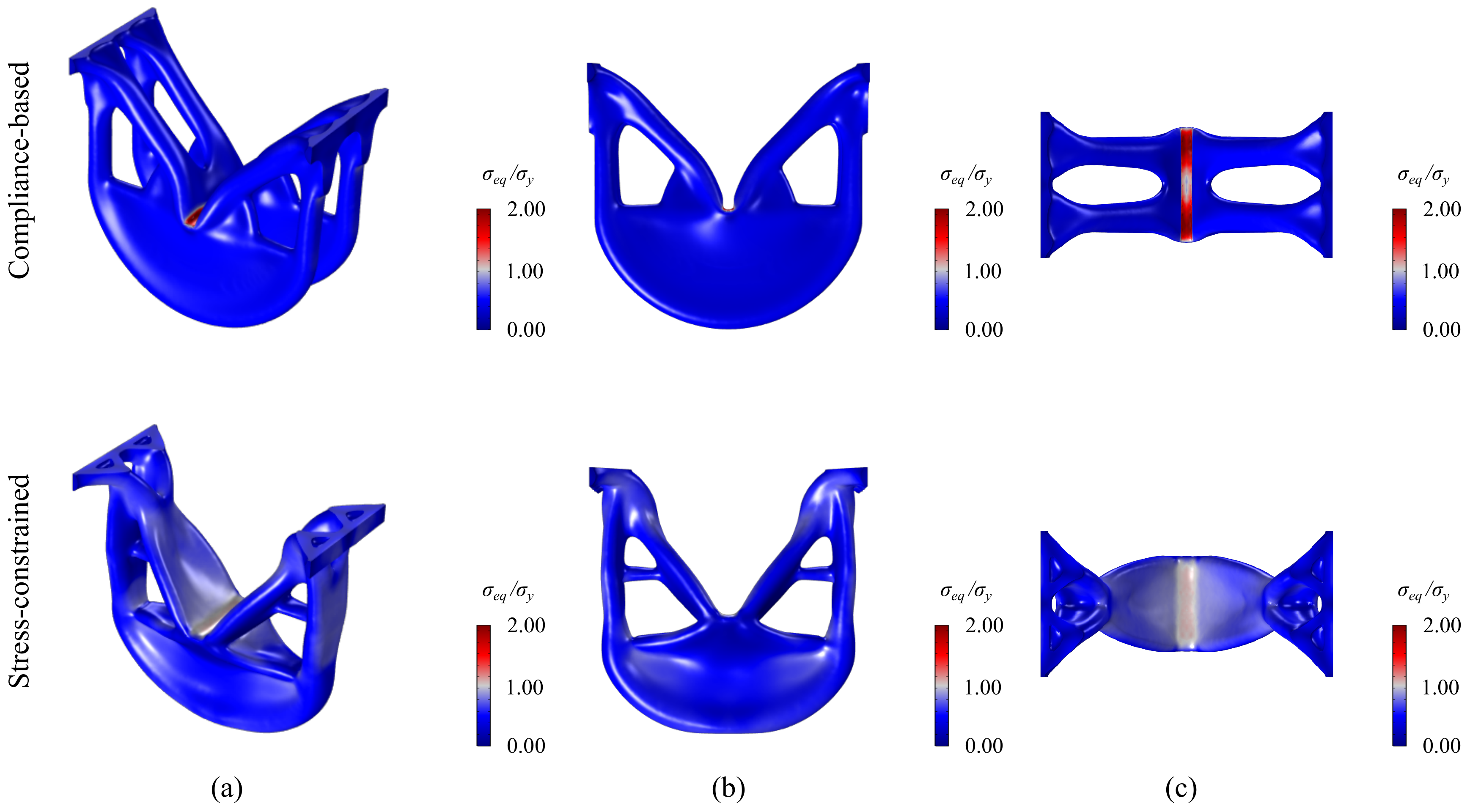}
\caption{Crack problem. Intermediate topologies for the COMSOL model obtained with compliance-based (first row) and stress-constrained (second row) robust formulations, from three different points of view: (a); (b); and (c).}
\label{C1_stress_flex_COMSOL_full}
\end{figure}

Figure \ref{C1_stress_flex_COMSOL_full} shows intermediate designs for compliance-based (first row) and stress-constrained (second row) formulations, from three different points of view. The full body-fitted (COMSOL) models are shown, with respective normalized von Mises stresses. Table \ref{Tabela_compara_stress_compliance} shows the compliance values and maximum von Mises equivalent stresses. The relative differences are computed by taking the stress-constrained case as reference. It is observed that the compliance-based result presents a structural compliance $27.79 \%$ lower, and a maximum von Mises stress $99.24 \%$ higher than the stress-constrained result. This is not surprising, since stress constraints are not taken into account in the compliance formulation, so that the optimizer does not see the need to promote a more rounded corner at the crack region in order to alleviate the stress concentration.

\begin{table}[ht!]
\centering
\caption{Structural compliance, maximum von Mises stress and maximum stress constraint violation for compliance-based and stress-constrained intermediate COMSOL models.}
\begin{tabular}{lccc}
\hline
Robust formulation & Compliance & $\sigma_{max}^{fitted}$ & $\frac{\sigma_{max}^{fitted}}{\sigma_{y}}-1$ \\
\hline
Compliance-based   & $263.68$ & $107.59$ & $115.18 \%$ \\
Stress-constrained & $365.15$ & $54.00$  & $8.00   \%$ \\
\hline
\end{tabular}
\label{Tabela_compara_stress_compliance}
\end{table}

\section{Concluding remarks}\label{s5}

This work has addressed truly large scale three-dimensional topology optimization problems under stress constraints and manufacturing uncertainty. The standard three-field density projection approach was generalized to a multi-field approach using a double filter. Several numerical examples were solved, and the results were post-processed with body-fitted finite element models using commercial software. The main conclusions are:
\begin{enumerate}
\item The employed solution procedure, based on the augmented Lagrangian method, is able to handle extremely large problems, with hundreds of millions of stress constraints.
\item The robust single filter procedure does not ensure identical eroded, intermediate and dilated topologies. It was observed that, in such cases, the intermediate topology is not robust with respect to uniform boundary variations, even though eroded and dilated designs satisfy the stress constraints. The proposed robust double filter approach, on the other hand, has achieved identical topologies and manufacturing tolerant results in all cases.
\item Good mesh independence was observed. Very small differences in shape and volume fraction were observed for medium and large mesh sizes.
\item The study on the influence of filter radius demonstrated that more than three fields of relative densities may be necessary to ensure manufacturing tolerance when combining small filter radius and large $\eta$ range. Moreover, an excellent agreement between expected and obtained tolerance ranges was observed.
\item Post-processing the results with body-fitted models demonstrated good stress accuracy of the voxel-based (fixed grid) models. Moreover, the body-fitted post-processing scheme demonstrated that the obtained optimized topologies are truly robust with respect to uniform boundary variation.
\item The proposed approach is general, having been successfully used to achieve several three-dimensional manufacturing tolerant designs that satisfy the stress constraints.
\end{enumerate}

Possibilities for future work include application to compliant mechanism design, topology design under dynamic response, and handling thermal stresses in topology optimization for additive manufacturing. Another possibility is investigation in the field of nonlinear mechanics.

\section*{Acknowledgments}
G. A. da Silva and A. T. Beck kindly acknowledge financial support of this research project by the agencies FAPESP (S\~ao Paulo Research Foundation), grant numbers 2018/16701-1 and 2019/08654-6, and CNPq (National Council for Research and Development), grant number 306373/2016-5. This study was financed in part by the Coordination for the Improvement of Higher Education Personnel - Brazil (CAPES) - Finance Code 001. N. Aage and O. Sigmund were supported by the Villum Investigator Project InnoTop funded by the Villum Foundation. Fruitful discussions with members of the DTU TopOpt-group are also gratefully acknowledged.

\subsection*{Conflict of interest}

The authors declare no potential conflict of interests.

\appendix

\section{\texorpdfstring{$\beta_{lim}$}{Lg} for PDE-based filter}\label{a1}

In Da Silva et al. \cite{Artigo6}, $\beta_{lim}$ is defined for the classical filter with linear hat function as an upper bound to $\beta$, Equation \eqref{projecao_Heaviside}. It ensures a smooth transition boundary, between solid and void phases, of length no less than $l_{e}$ (element size). In the current manuscript, however, the PDE filter is employed. This section presents the step-by-step procedure by Da Silva et al. \cite{Artigo6}, but for the PDE filter instead of the linear one. As a result, $\beta_{lim}^{PDE}$ is achieved.

\begin{figure}[ht!]
\centering
\includegraphics[width=0.99\textwidth]{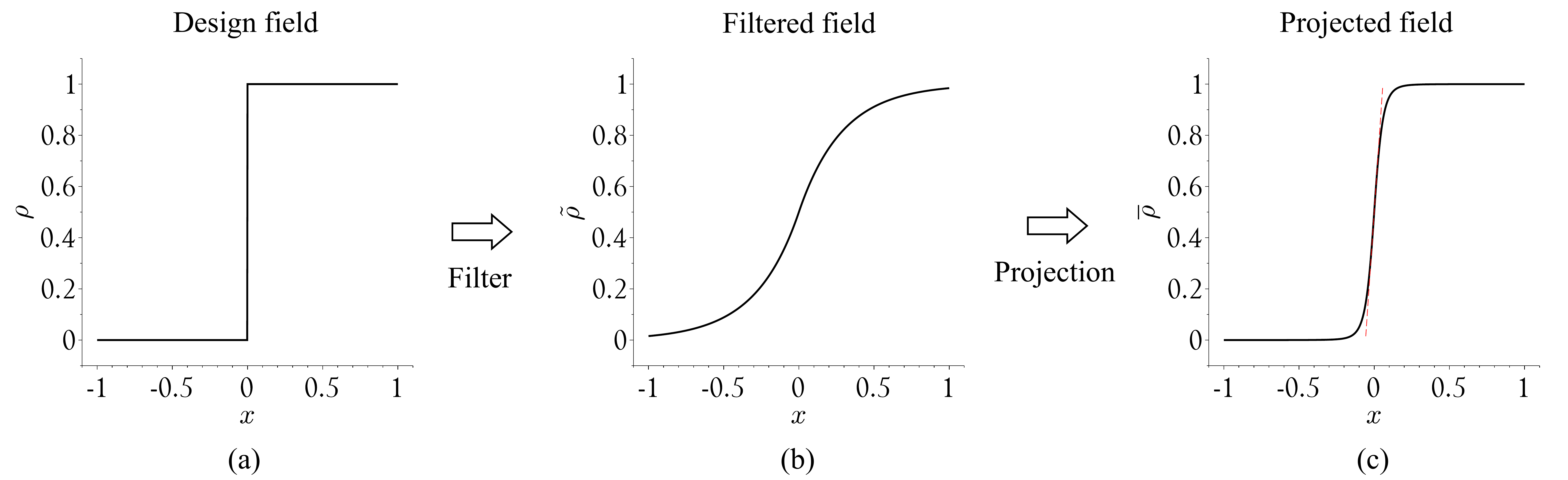}
\caption{(a) Design field; (b) Filtered field, for $R = 1$; (c) Projected field, for $\eta = 0.5$ and $\beta = 10$. Dashed line in projected field is tangent to the curve in $x = 0$.}
\label{vp_vf_vp}
\end{figure}

Following Da Silva et al. \cite{Artigo6}, we assume the one-dimensional design field of Figure \ref{vp_vf_vp} (a), given by the Heaviside step function
\begin{equation}
\rho \left( x \right) = \left\lbrace \begin{array}{lll}
0 & & \text{if } x < 0 \\
1 & & \text{if } x \geqslant 0 \\
\end{array} \right. \\.\label{rho}
\end{equation}

The solution of Equation \eqref{filtro_PDE} can be expressed as a convolution integral, such as performed for the classical filter with linear hat function \citep{Lazarov_PDE}. For one-dimensional problems, one can write
\begin{equation}
\tilde{\rho} = \frac{\int_{-\infty}^{\infty} w \left( x,x_{c} \right) \rho \left( x \right) \; \text{d}x_{c}}{\int_{-\infty}^{\infty} w \left( x,x_{c} \right) \; \text{d}x_{c}},\label{rhoTilde_1}
\end{equation}
where $w \left( x,x_{c} \right)$ is the Green's function centered at $x_{c}$, given by
\begin{equation}
w \left( x,x_{c} \right) = \frac{1}{2 R_{PDE}} \exp \left( - \frac{\left\vert x_{c} - x \right\vert}{R_{PDE}} \right).\label{Greens_function}
\end{equation}

Solving Equation \eqref{rhoTilde_1} using the Green's function, we get
\begin{equation}
\tilde{\rho} \left( x \right) = \left\lbrace \begin{array}{lll}
\frac{1}{2} \exp \left( \frac{2 \sqrt{3} x}{R} \right) & & \text{if } x < 0 \\
1 - \frac{1}{2} \exp \left( -\frac{2 \sqrt{3} x}{R} \right) & & \text{if } x \geqslant 0 \\
\end{array} \right. \\ , \label{rhoTilde_2}
\end{equation}
which is shown in Figure \ref{vp_vf_vp} (b), for $R = 1$.

Substituting Equation \eqref{rhoTilde_2} in Equation \eqref{projecao_Heaviside} gives the projected field, which is illustrated in Figure \ref{vp_vf_vp} (c) for $\eta = 0.5$ and $\beta = 10$. The derivative of the resulting equation with respect to $x$ is the incline of the projected field by definition, which should be limited to ensure a smooth transition boundary of prescribed length. Through trivial algebra, one can demonstrate that the largest incline occurs for $x = 0$ and $\eta = 0.5$, and it is given by
\begin{equation}
\tan(\alpha_{max}) = \frac{0.5 \sqrt{3} \beta}{R \tanh(0.5 \beta)},\label{tan_alpha_max}
\end{equation}
illustrated by the dashed line in Figure \ref{vp_vf_vp} (c).

For large values of $\beta$, one can use Equation \eqref{tan_alpha_max} to write
\begin{equation}
\lim_{\beta\rightarrow\infty} \frac{\tan(\alpha_{max})}{\beta} = \frac{\sqrt{3}}{2 R}.\label{tan_alpha_max_2}
\end{equation}

Based on Equation \eqref{tan_alpha_max_2}, the largest incline for a large value of $\beta$ can be defined as
\begin{equation}
\tan(\alpha_{Max}) = \frac{\beta \sqrt{3}}{2 R}.\label{tan_alpha_max_3}
\end{equation}

Similarly, one can define a limiting incline based on the size $l_{e}$, given by
\begin{equation}
\tan(\alpha_{lim}) = \frac{1}{l_{e}}.\label{tan_alpha_lim}
\end{equation}

In order to find $\beta_{lim}^{PDE}$, we match Equations \eqref{tan_alpha_max_3} and \eqref{tan_alpha_lim}, which gives
\begin{equation}
\beta_{lim}^{PDE} = \frac{2 R}{l_{e}\sqrt{3}}.\label{beta_lim}
\end{equation}

\section{Length scale and manufacturing tolerance for PDE-based filter}\label{a2}

In order to achieve the expected minimum length scale and minimum manufacturing tolerance for the PDE filter, we follow the step-by-step procedure by Wang et al. \cite{Wang2011} and Qian and Sigmund \cite{Qian_Sigmund_2013}, but for the Green's function instead of the linear one.

\begin{figure}[ht!]
\centering
\includegraphics[width=0.8\textwidth]{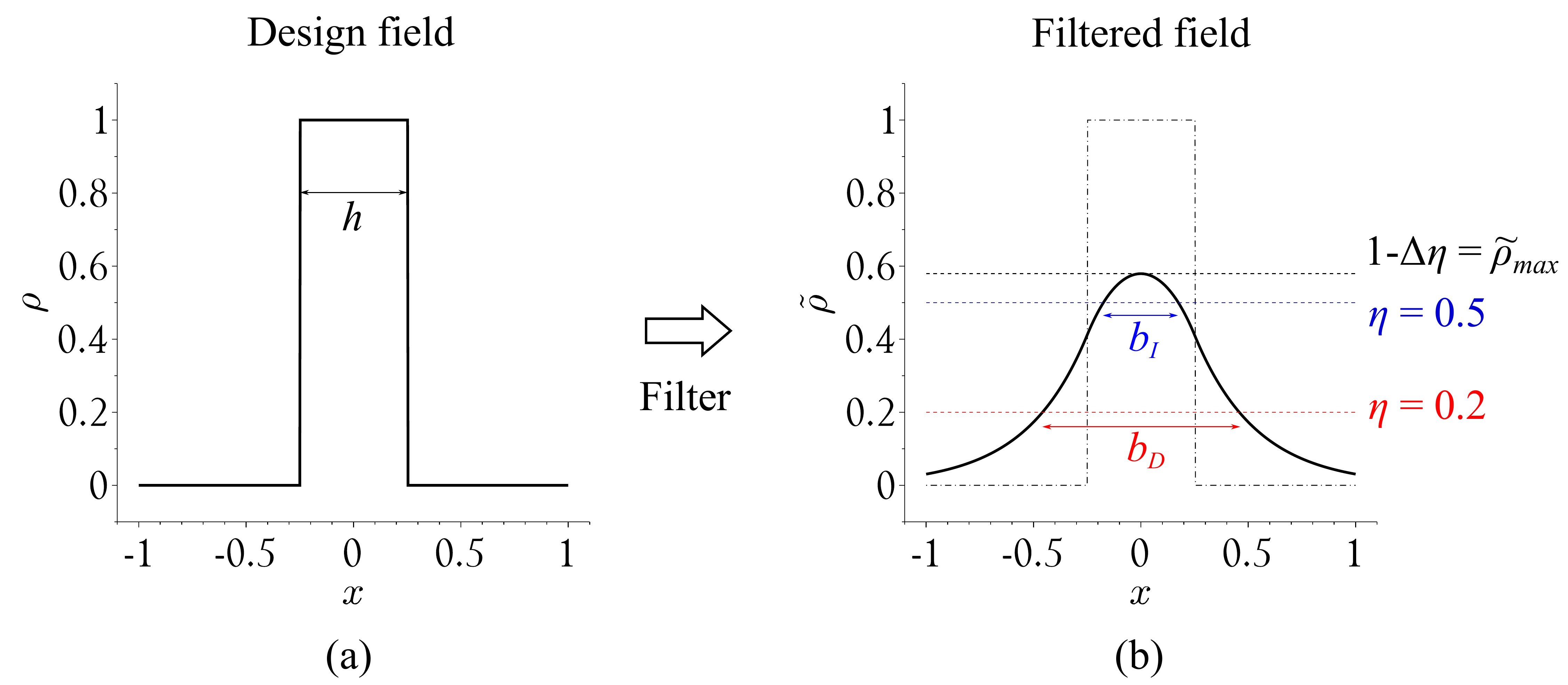}
\caption{(a) Design field (solid line), for $h = 0.5$; (b) Filtered field (solid line), for $R = 1$. In (b), dash-dotted line represents the design field, and horizontal dashed lines represent specific $\eta$ values: $1 - \Delta \eta$, $0.5$ and $0.2$; from top to bottom, respectively.}
\label{vp_vf_h}
\end{figure}

We assume the following one-dimensional design field
\begin{equation}
\rho \left( x \right) = \left\lbrace \begin{array}{lll}
0 & & \text{if } x < -h/2 \\
1 & & \text{if } -h/2 \leqslant x < h/2 \\
0 & & \text{if } x \geqslant h/2 \\
\end{array} \right. \\,\label{rho_h}
\end{equation}
represented in Figure \ref{vp_vf_h} (a), for $h = 0.5$. Substituting Equation \eqref{rho_h} in Equation \eqref{rhoTilde_1} gives
\begin{equation}
\tilde{\rho} \left( x \right) = \left\lbrace \begin{array}{lll}
\frac{1}{2} \left[ \exp \left( \frac{\sqrt{3}}{R} \left( 2x + h \right) \right) - \exp \left( \frac{\sqrt{3}}{R} \left( 2x - h \right) \right) \right] & & \text{if } x < -h/2 \\
1 - \frac{1}{2} \left[ \exp \left( \frac{\sqrt{3}}{R} \left( 2x - h \right) \right) + \exp \left( -\frac{\sqrt{3}}{R} \left( 2x + h \right) \right) \right] & & \text{if } -h/2 \leqslant x < h/2 \\
\frac{1}{2} \left[ \exp \left( \frac{\sqrt{3}}{R} \left( h - 2x \right) \right) - \exp \left( -\frac{\sqrt{3}}{R} \left( 2x + h \right) \right) \right] & & \text{if } x \geqslant h/2 \\
\end{array} \right. \\,\label{rhoTilde_h}
\end{equation}
which represents the filtered field, illustrated in Figure \ref{vp_vf_h} (b), for $R = 1$.

\begin{figure}[ht!]
\centering
\includegraphics[width=0.4\textwidth]{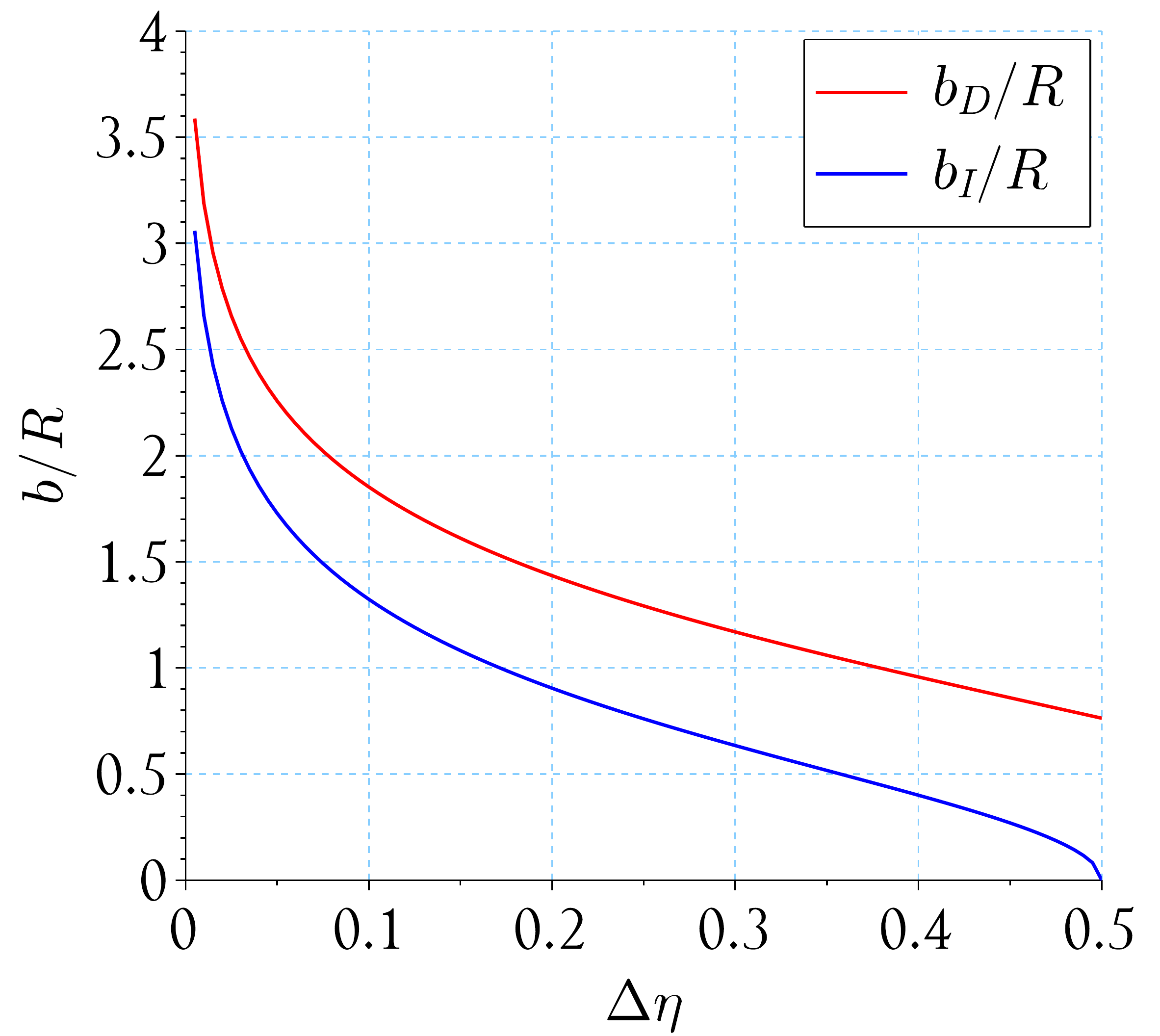}
\caption{Minimum length scales for intermediate, $b_{I}$, and dilated, $b_{D}$, topologies, for different $\Delta \eta$ values.}
\label{bI_bD}
\end{figure}

The minimum length scale on the solid phase of the intermediate design, given by $b_{I}$, is computed by assuming identical eroded and intermediate topologies. When the eroded topology reduces to a point, related to $\eta_{max} = 1 - \Delta \eta$ in Figure \ref{vp_vf_h} (b), one can compute $b_{I}$ as the distance between intersections of $\tilde{\rho} \left( x \right)$, Equation \eqref{rhoTilde_h}, with the horizontal dashed line for $\eta = 0.5$ (which represents the intermediate design in this paper). Through trivial algebra, one can demonstrate that the minimum length scale on the solid phase of the intermediate topology can be computed as
\begin{equation}
\frac{b_{I}}{R} = \frac{1}{2\sqrt{3}} \ln \left( \frac{1 + \sqrt{1-4 \left( \Delta \eta \right)^2}}{1 - \sqrt{1-4 \left( \Delta \eta \right)^2}} \right), \label{b_I}
\end{equation}
where $\Delta \eta$ controls the $\eta$ value of the eroded topology. Figure \ref{bI_bD} illustrates the ratio $b_{I}/R$ for $\Delta \eta \in \left(0,0.5\right]$. The minimum length scale $b_{I}$ represents the diameter of a circular length scale in 2D, or spherical in 3D.

The minimum length scale on the void phase is computed in a similar way \citep{Wang2011}. Instead of using the design field from Equation \eqref{rho_h}, one has to consider the complementary situation, i.e., considering the distance $h$ composed of void, and the remainder of solid. In this paper, since $\eta_{max} - \eta_{i} = \eta_{i} - \eta_{min}$, where $\eta_{min}$ and $\eta_{max}$ are the minimum and maximum $\eta$ values considered during optimization, respectively, and $\eta_{i} = 0.5$ refers to the intermediate design, we have the same minimum length scale for solid and void phases, and it is given by Equation \eqref{b_I}.

In order to compute the minimum manufacturing tolerance between intermediate and dilated topologies, one has to compute the minimum length scale on the solid phase of the dilated topology. Following the same procedure as before for a dilated design defined for $\eta = 0.2$, for instance, we have
\begin{equation}
\frac{b_{D}}{R} = \frac{1}{\sqrt{3}} \ln \left( \frac{1 - \left( \Delta \eta \right)^2}{0.4 \Delta \eta} \right), \label{b_D}
\end{equation}
which is illustrated in Figure \ref{bI_bD}.

The minimum manufacturing tolerance is then given by a circular (in 2D) or spherical (in 3D) length scale with diameter equal to $\left(b_{D} - b_{I} \right)/2$. The same step-by-step procedure can be employed to compute the minimum manufacturing tolerance between eroded and intermediate topologies; however, due to the symmetry conditions regarding the $\eta$ range employed in this paper, the same equation can be employed for this situation, which gives a minimum distance between eroded and dilated topology equal to twice the minimum manufacturing tolerance, i.e., $b_{D} - b_{I}$ for $\Delta \eta = 0.2$.

\section{Sensitivity analysis}\label{a3}

Sensitivity analysis for the stress-constrained volume minimization problem considering the three-field density projection approach is shown in Da Silva et al. \cite{Artigo6} in detail. The extension to address $N_{j}$ fields of relative densities is straightforward, and the step-by-step procedure is not presented herein. Instead, we just show how the augmented Lagrangian may be written to facilitate use of the adjoint method in this case, as well as the adjoint problem and final derivative.

The augmented Lagrangian function, Equation \eqref{LA_robusto}, is rewritten as
\begin{equation}
L = V^{(1)} + \sum_{j = 1}^{N_{j}} L^{(j)},\label{LA_robusto_2}
\end{equation}
where $V^{(1)}$ is the normalized volume of the dilated structure, written as
\begin{equation}
V^{(1)} = \frac{N_{k} \sum_{e=1}^{N_{e}} V_{e} \overline{\overline{\rho}}_{e}^{(1)}}{\sum_{e=1}^{N_{e}}V_{e}},\label{volume_normalizado}
\end{equation}
and $L^{(j)}$ is the term associated with the stress constraints, given by
\begin{equation}
L^{(j)} = \frac{r}{2}\sum_{k=1}^{N_{k}} \left\langle \frac{\mu_{k}^{(j)}}{r} + \frac{\sigma_{eq}^{(k)}\left(\overline{\overline{\bm{\rho}}}^{(j)}\right)}{\sigma_{y}} - 1 \right\rangle^{2} + \bm{\lambda}_{(j)}^{T}\left( \mathbf{K}\left(\overline{\overline{\bm{\rho}}}^{(j)}\right)\mathbf{U}\left(\overline{\overline{\bm{\rho}}}^{(j)}\right) - \mathbf{F} \right),\label{LA_robusto_3}
\end{equation}
where $\bm{\lambda}_{(j)}$ is the $j$-th adjoint vector, which is arbitrary, since $\mathbf{K}\left(\overline{\overline{\bm{\rho}}}^{(j)}\right)\mathbf{U}\left(\overline{\overline{\bm{\rho}}}^{(j)}\right) - \mathbf{F} = \mathbf{0}$.

The derivative of the normalized dilated volume, Equation \eqref{volume_normalizado}, is straightforward, and is given by
\begin{equation}
\frac{\partial V^{(1)}}{\partial \overline{\overline{\rho}}_{n}^{(1)}} = \frac{N_{k} V_{n}}{\sum_{e=1}^{N_{e}}V_{e}}.\label{derivada_volume_normalizado}
\end{equation}

The derivative of $L^{(j)}$, Equation \eqref{LA_robusto_3}, is shown in detail by Da Silva et al. \cite{Artigo6}, and is given by
\begin{equation}
\frac{\partial L^{(j)}}{\partial \overline{\overline{\rho}}_{n}^{(j)}} = h_{n}^{(j)} \frac{\partial f_{\sigma} \left( \overline{\overline{\rho}}_{n}^{(j)} \right)}{\partial \overline{\overline{\rho}}_{n}^{(j)}} \hat{\sigma}_{eq}^{(n)}\left(\overline{\overline{\bm{\rho}}}^{(j)}\right) + \bm{\lambda}_{(j,n)}^{T} \frac{\partial \mathbf{k}_{n}\left(\overline{\overline{\rho}}^{(j)}_{n}\right)}{\partial \overline{\overline{\rho}}_{n}^{(j)}} \mathbf{u}_{n}\left(\overline{\overline{\bm{\rho}}}^{(j)}\right),\label{LA_dif_3}
\end{equation}
where
\begin{equation}
h_{n}^{(j)} = \left\langle \mu_{n}^{(j)} + r \left( \frac{\sigma_{eq}^{(n)}\left(\overline{\overline{\bm{\rho}}}^{(j)}\right)}{\sigma_{y}} - 1 \right) \right\rangle \frac{1}{\sigma_{y}},\label{hk}
\end{equation}
and $\bm{\lambda}_{(j,n)} = \mathbf{H}_{n} \bm{\lambda}_{(j)}$ is the local adjoint vector, obtained by use of the localization operator $\mathbf{H}_{n}$ \citep{Bathe}.

The $j$-th adjoint vector, $\bm{\lambda}_{(j)}$, is the solution of the following system of linear equations
\begin{equation}
\mathbf{K}\left(\overline{\overline{\bm{\rho}}}^{(j)}\right) \bm{\lambda}_{(j)} = - \sum_{k=1}^{N_{k}} h_{k}^{(j)} \frac{f_{\sigma} \left( \overline{\overline{\rho}}_{k}^{(j)} \right)}{\hat{\sigma}_{eq}^{(k)}\left(\overline{\overline{\bm{\rho}}}^{(j)}\right)} \mathbf{H}_{k}^{T} \mathbf{a}_{k}^{(j)},\label{eq_adjunta}
\end{equation}
where
\begin{equation}
\mathbf{a}_{k}^{(j)} = \mathbf{B}_{k}^{T} \mathbf{C}^{0} \mathbf{M} \mathbf{C}^{0} \mathbf{B}_{k} \mathbf{u}_{k}\left(\overline{\overline{\bm{\rho}}}^{(j)}\right).\label{ak}
\end{equation}

After obtaining the derivatives of $V^{(1)}$ and $L^{(j)}$ with respect to the physical relative densities, the chain rule is employed, as shown in Christiansen et al. \cite{Christiansen2015} for the double filter procedure, to achieve the derivatives with respect to a design variable $\rho_{m}$.

Note that $N_{j}$ adjoint problems, Equation \eqref{eq_adjunta}, are solved to evaluate Equation \eqref{LA_dif_3} for $j = 1,2,...,N_{j}$, i.e., one adjoint problem per physical density field.

\section{Steepest descent method with move limits}\label{a4}

To solve the optimization subproblems, Equation \eqref{Problema_otm_4}, we use a simplified version of the modified steepest descent method proposed by Da Silva et al. \cite{Artigo3}. Given the gradient of the augmented Lagrangian function at iteration $b$ (Appendix \ref{a3}), written as $\left.\nabla_{\bm{\rho}} L \; \right\vert_{\bm{\rho} = \bm{\rho}^{(b)}}$, we find the design variables at iteration $b+1$, given by $\bm{\rho}^{(b+1)}$, through the following procedure
\begin{enumerate}
\item Given $\bm{\rho}^{(b)}$, compute the steepest descent direction, as $\mathbf{S} = -\left.\nabla_{\bm{\rho}} L \; \right\vert_{\bm{\rho} = \bm{\rho}^{(b)}}$
\item Reset the gradient contributions at the bound constraints, as follows
\begin{equation}
S_{e} = \left\lbrace \begin{array}{lll}
0     &  & \text{if } \rho_{e} = 1 \text{ and } S_{e} > 0 \\
0     &  & \text{if } \rho_{e} = 0 \text{ and } S_{e} < 0 \\
S_{e} &  & \text{othrewise } \\
\end{array} \right.
\end{equation}
\item Normalize $\mathbf{S}$ by its maximum (absolute) value, as $\mathbf{D} = \frac{\mathbf{S}}{\max \left\vert \mathbf{S} \right\vert}$
\item Set move limits based on two previous iterations, by using the auxiliary variable $d_{e} = \left( \rho_{e}^{(b)} - \rho_{e}^{(b-1)} \right) \times \left( \rho_{e}^{(b-1)} - \rho_{e}^{(b-2)} \right)$, as follows
\begin{equation}
\begin{array}{l}

\delta_{e} = \left\lbrace \begin{array}{lll}
0.7 \; \delta_{e} & & \text{if } d_{e} < 0 \\
1.1 \; \delta_{e} & & \text{if } d_{e} > 0 \\
\end{array} \right. \\

\\

\delta_{e} = \left\lbrace \begin{array}{lll}
0.1   & & \text{if } \delta_{e} > 0.1 \\
0.001 & & \text{if } \delta_{e} < 0.001 \\
\end{array} \right. \\

\\

\rho_{e}^{inf} = \rho_{e}^{(b)} - \delta_{e} \\
\rho_{e}^{sup} = \rho_{e}^{(b)} + \delta_{e}

\end{array}
\end{equation}
\item Compute $\bm{\rho}^{(b+1)}$ by using a unitary step length $\Psi = 1$, with the following update procedure
\begin{equation}
\rho_{e}^{(b+1)} = \left\lbrace \begin{array}{lll}
\max \left( \rho_{e}^{inf} , 0 \right) & & \text{if } \rho_{e}^{(b)} + \Psi D_{e} \leqslant \max \left( \rho_{e}^{inf} , 0 \right) \\
\min \left( \rho_{e}^{sup} , 1 \right) & & \text{if } \rho_{e}^{(b)} + \Psi D_{e} \geqslant \min \left( \rho_{e}^{sup} , 1 \right) \\
\rho_{e}^{(b)} + \Psi D_{e} & & \text{otherwise} \\
\end{array} \right. ,
\end{equation}
\end{enumerate}
where $\rho_{e}^{inf}$ and $\rho_{e}^{sup}$ are the lower and upper move limits associated with the $e$-th design variable, respectively, and $\delta_{e}$ is an auxiliary variable employed to compute them. The optimization procedure is started with maximum range of move limits, i.e. $\delta_{e} = 0.1$, and these are updated from the third iteration. The main difference between the employed procedure and the one originally proposed by Da Silva et al. \cite{Artigo3} is the step length $\Psi$, which is unitary in the current implementation. One could employ the backtracking algorithm proposed by Da Silva et al. \cite{Artigo3} instead, to ensure the minimization of the augmented Lagrangian function within the iterations of a given subproblem; however, we observed good convergence in all cases analyzed in this paper by using the unitary step length and hence, we preferred to keep this modification for simplicity.

\bibliography{bibliografia}

\end{document}